\documentclass[aps,prx,twocolumn,longbibliography,10pt]{revtex4-2}

\usepackage{amsmath}
\usepackage{amssymb}
\usepackage{graphicx}
\usepackage[colorlinks=true]{hyperref}
\usepackage{bm}
\usepackage{subfigure}
\usepackage{xcolor}
\usepackage{enumitem}
\usepackage{array}

\newcommand{\Z}{\mathbb{Z}}
\newcommand{\EE}{\mathcal{E}}
\newcommand{\CC}{\mathcal{C}}
\newcommand{\ZZ}{\mathcal{Z}}
\newcommand{\HH}{\mathcal{H}}
\newcommand{\bfX}{\mathbf{X}}
\newcommand{\bfZ}{\mathbf{Z}}
\newcommand{\llangle}{\langle \! \langle}
\newcommand{\rrangle}{\rangle \! \rangle}
\newcommand{\bbE}{\mathbb{E}}
\newcommand{\bfx}{\mathbf{x}}
\newcommand{\bfz}{\mathbf{z}}
\newcommand{\bft}{\mathbf{t}}
\newcommand{\bfs}{\mathbf{s}}
\newcommand{\bftau}{\bm{\tau}}
\newcommand{\bfsig}{\bm{\sigma}}
\newcommand{\Tr}{\mathrm{Tr}}
\newcommand{\fkg}{\mathfrak{g}}
\newcommand{\LL}{\mathcal{L}}
\newcommand{\DD}{\mathcal{D}}
\newcommand{\TT}{\mathcal{T}}
\newcommand{\OO}{\mathcal{O}}
\newcommand{\SSS}{\mathcal{S}}
\newcommand{\JJ}{\mathcal{J}}
\newcommand{\KK}{\mathcal{K}}
\newcommand{\QQ}{\mathcal{Q}}
\newcommand{\RR}{\mathcal{R}}
\newcommand{\sech}{\mathrm{sech}}

\definecolor{darkblue}{rgb}{0,0,1}

\begin{document}

\title{Mixed-State Topological Order under Coherent Noise}

\author{Seunghun Lee}
\affiliation{Department of Physics, Korea Advanced Institute of Science and Technology, Daejeon 34141, Republic of Korea}
\author{Eun-Gook Moon}
\thanks{egmoon@kaist.ac.kr}
\affiliation{Department of Physics, Korea Advanced Institute of Science and Technology, Daejeon 34141, Republic of Korea}

\begin{abstract}
    Mixed-state phases of matter under local decoherence have recently garnered significant attention due to the ubiquitous presence of noise in current quantum processors. One of the key issues is understanding how topological quantum memory is affected by realistic coherent noise, such as random rotation noise and amplitude damping noise. In this work, we investigate the intrinsic error threshold of the two-dimensional toric code, a paradigmatic topological quantum memory, under these coherent noise by employing both analytical and numerical methods based on the doubled Hilbert space formalism. A connection between the mixed-state phase of the decohered toric code and a non-Hermitian Ashkin-Teller-type statistical mechanics model is established, and the mixed-state phase diagrams under the coherent noise are obtained. We find remarkable stability of mixed-state topological order under random rotation noise with axes near the $Y$-axis of qubits. We also identify intriguing extended critical regions at the phase boundaries, highlighting a connection with non-Hermitian physics. We argue that these phase boundaries provide upper bounds for the intrinsic error threshold, beyond which quantum error correction becomes impossible. We complement these findings by estimating the error thresholds for random rotation noise under standard quantum error correction, thereby providing lower bounds on the intrinsic error threshold.
\end{abstract}

\maketitle

\section{Introduction}

Topologically ordered phases of matter, which go beyond the conventional Landau paradigm, have been extensively investigated in the context of pure states~\cite{wen2004quantum,sachdev2023quantum}. These quantum many-body states exhibit exotic properties such as long-range entanglement, robust ground-state degeneracy, and localized excitations with nontrivial exchange statistics. These features are resilient against local perturbations, making topological phases a promising platform for quantum error correction (QEC) and fault-tolerant quantum computing~\cite{kitaev2003fault,dennis2002topological}. The toric code, epitomizing $\Z_2$ topological order, serves as a prime example of a topological QEC code that can encode two logical qubits within its ground-state subspace~\cite{kitaev2003fault}. 

Recently, significant progress has been made in realizing topologically ordered states on programmable quantum simulators~\cite{semeghini2021probing,satzinger2021realizing,foss2023experimental,iqbal2024topological,iqbal2024non,xu2024non,minev2025realizing}. However, the inevitable presence of noise in current noisy intermediate-scale quantum (NISQ) devices~\cite{preskill2018quantum} renders the prepared quantum states as mixed states. Consequently, the mixed-state phase of matter has attracted considerable attention as a topic of both fundamental interest and practical importance~\cite{lee2025symmetry,bao2023mixed,lee2023quantum,fan2024diagnostics,lee2025exact,niwa2025coherent,de2022symmetry,ma2023average,coser2019classification,sang2024mixed,sang2025stability,zou2023channeling,lu2023mixed,guo2024two,ma2025topological,ma2025symmetry,chen2023symmetry,chen2024separability,chen2024unconventional,myerson2025decoherence,wang2025intrinsic,su2024higher,su2024conformal,lyons2024,su2024tapestry,li2025replica,zhang2025quantum,lu2024disentangling,sohal2025noisy,ellison2025toward,lessa2025mixed,lessa2025strong,sala2024spontaneous,hauser2024information,eckstein2024robust,shah2024instability,sun2025holographic,you2024intrinsic,sala2025stability,guo2025locally,xue2024tensor,kim2024error,kim2024persistent,gu2024spontaneous,huang2025hydrodynamics,zhang2025strong,negari2024spacetime}. In particular, examining the mixed-state topological order can offer valuable insights into the intrinsic QEC properties of quantum memories under decoherence. For example, studies have shown that the toric code subject to incoherent bit-flip or phase-flip errors undergoes a mixed-state phase transition at a critical noise strength~\cite{fan2024diagnostics,bao2023mixed,lee2023quantum}, which coincides with the error threshold of the toric code with the optimal decoder~\cite{dennis2002topological}. 

Previous studies on mixed-state topological order have mostly focused on incoherent local noise, where Pauli errors stochastically affect qubits. However, realistic quantum processors encounter \emph{coherent} errors, which create a coherent superposition of different error states and give non-unique syndromes. Typical coherent errors include systematic unitary rotations caused by imperfect gate operations~\cite{kueng2016comparing,greenbaum2017modeling,bravyi2018correcting,venn2023coherent,tomita2014low,hakkaku2021sampling,pataki2024coherent,marton2023coherent,venn2020error} and amplitude damping due to spontaneous emission~\cite{nielsen2001quantum,chirolli2008decoherence,eczoo_ampdamp}. Therefore, it is crucial to study mixed-state topological order under coherent noise and understand their impact on quantum memories.

\begin{figure*}[t]
    \centering
    \includegraphics[width=2\columnwidth]{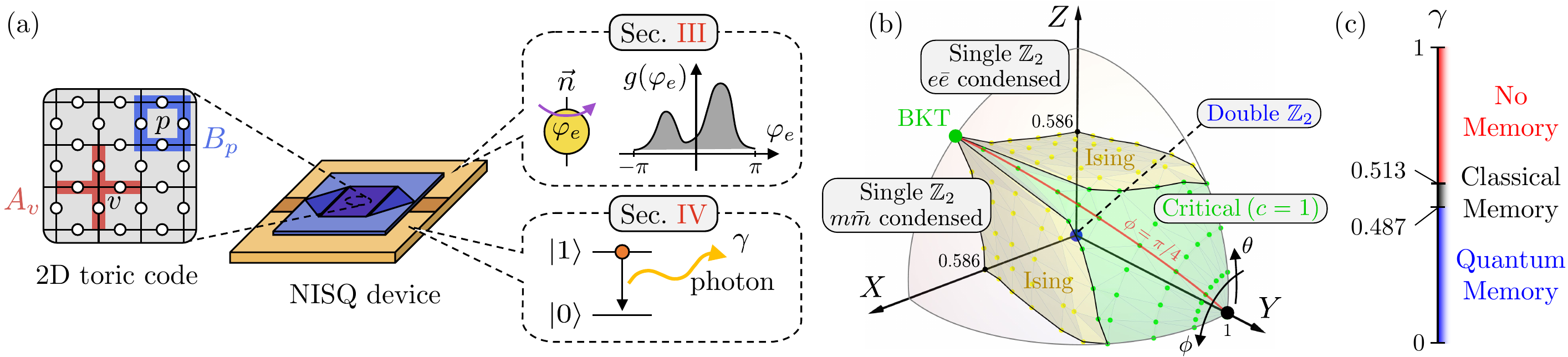}
    \caption{(a) Schematic of the coherent noise models under consideration. The two-dimensional toric code [see Eq.~\eqref{TC}] is subject to two types of coherent noise relevant to current noisy intermediate-scale quantum (NISQ) devices: (i) random rotation noise, where each qubit is rotated around the $\vec{n}$ axis by an angle $\varphi_e$ according to distribution $g(\varphi_e)$ (see Sec.~\ref{Sec:RandRotNoise}), and (ii) amplitude damping noise, where each qubit decays to the ground state via spontaneous emission with probability $\gamma$ (see Sec.~\ref{Sec:AmpDampNoise}). (b) Phase diagram of the decohered toric code under random rotation noise. Radial direction and distance represent the rotation axis $\vec{n}$ and a parameter $R \in [0, 1]$ that encodes information about the angle distribution $g(\varphi_e)$ [see Eq.~\eqref{R}], respectively. The system possesses quantum (classical) memory within (outside) the region enclosed by yellow and green phase boundaries. Two yellow boundaries belong to 2D Ising universality ($c = 1/2$) and the green boundary with $R = 1$ forms an extended critical region with $c = 1$. A pure toric code is marked by the blue point located at the origin. (c) Phase diagram of the decohered toric code under amplitude damping noise. All phase diagrams illustrated here are based on the doubled Hilbert space formalism.}
    \label{fig:Setup_Summary}
\end{figure*}

In this work, we address these questions by studying the two-dimensional (2D) toric code under two types of local coherent noise: (i) random rotation noise, where each qubit experiences unitary rotation around the $\vec{n}$ axis by an angle $\varphi_e$ according to some angle distribution $g(\varphi_e)$, and (ii) amplitude damping noise with damping parameter $\gamma$ [see Fig.~\ref{fig:Setup_Summary}(a)]. We investigate the mixed-state phases of the decohered toric code under these noise models based on the doubled Hilbert space formalism~\cite{lee2025symmetry,bao2023mixed,lee2023quantum}. We do so by mapping the R\'enyi-2 versions of information-theoretic quantities, such as anyon condensation/confinement parameters~\cite{bao2023mixed,haegeman2015shadows,duivenvoorden2017entanglement} and coherent information~\cite{schumacher1996quantum,lloyd1997capacity}, to observables in effective statistical mechanics (stat-mech) models. We show that these stat-mech models are the Ashkin-Teller-type models (with possibly anisotropic and imaginary coupling constants) for the considered coherent noise. 

Specifically, we prove that the mixed-state phase of the decohered toric code under random rotation noise depends on the angle distribution $g(\varphi_e)$ solely through a single parameter $R$ [defined in Eq.~\eqref{R}]. With this simplification, we map the mixed-state phase diagram under random rotation noise for any rotation axis $\vec{n}$ and arbitrary distribution $g(\varphi_e)$, using a combination of analytical results and tensor network simulations. Notably, we find that mixed-state topological order is highly resilient against random rotations with $\vec{n}$ near the $Y$-axis, where the phase boundary forms an extended critical region characterized by a central charge $c = 1$. Additionally, we identify an intriguing two-dimensional phase boundary with Ising criticality, extended by the non-Hermitian nature of the effective stat-mech model. We also complement the limitation of the doubled Hilbert space formalism by numerically estimating the error thresholds in standard QEC, which provide lower bounds for the mixed-state phase boundary in the replica limit.

For amplitude damping noise, we discover two successive phase transitions as the damping rate $\gamma$ increases, first degrading the quantum memory to classical memory, and then to no memory. Under physically reasonable assumptions, we argue that the mixed-state phase boundaries identified here establish upper bounds on the error threshold for coherent noise, within which QEC is feasible and beyond which QEC fails. The setup and the main results of this paper are illustrated in Fig.~\ref{fig:Setup_Summary}.    

The rest of the paper is structured as follows. Sec.~\ref{Sec:Background} briefly review the 2D toric code, the doubled Hilbert state formalism, and the considered information-theoretic diagnostics. Sec.~\ref{Sec:RandRotNoise} introduces random rotation noise and develops mapping to the stat-mech model. We discuss the general phase diagram and delineate the analytically tractable cases and the identified critical regions. Sec.~\ref{Sec:AmpDampNoise} presents the mixed-state phase diagram under the amplitude damping noise. Finally, we summarize our results and discuss open questions in Sec.~\ref{Sec:Discussion}.

\section{Background} \label{Sec:Background}

In this section, we provide a brief review of the 2D toric code and the doubled Hilbert space formalism. We then introduce two diagnostics used in subsequent sections to analyze mixed-state topological order under coherent noise: the anyon condensation/confinement parameters and the coherent information.

\subsection{2D Toric Code} \label{Sec:2DTC}

Consider an $L\times L$ square lattice with $N = 2L^2$ qubits on its edges. The Hamiltonian of the 2D toric code (TC) defined on this lattice is given by~\cite{kitaev2003fault}
\begin{align} \label{TC}
    H = -\sum_v A_v - \sum_p B_p,
\end{align}
where $A_v = \prod_{e \in \partial v} X_e$ and $B_p = \prod_{e\in \partial p} Z_e$ are the star and plaquette operators, respectively [see Fig.~\ref{fig:Setup_Summary}(a)]. The vertices and edges of the lattice are denoted by $v$ and $e$, respectively, and ($X$, $Y$, $Z$) are the usual Pauli operators. Since $A_v$ and $B_p$ mutually commute, the ground states $| \Psi_0 \rangle$ satisfy $A_v | \Psi_0 \rangle = B_p | \Psi_0 \rangle = | \Psi_0 \rangle$ for all $v$ and $p$. On a torus, the ground-state subspace is fourfold degenerate, which can be utilized as a code space encoding two logical qubits. 

We introduce an expression for TC ground states, which has been useful in studying TC under decoherence~\cite{chen2024separability, chen2024unconventional}. Suppose that qubits are also placed on the vertices of the square lattice. The cluster state defined on this lattice is given by~\cite{briegel2001persistent}
\begin{align}
    | \Psi_c \rangle \propto \prod_e (I + Z_e X_v X_{v'}) | \bfz_v = 1, \bfx_e = 1 \rangle,
\end{align}
where $v$ and $v'$ are vertices connected by edge $e = (v, v')$, $| \bfz_v = 1 \rangle$ is a tensor product of the eigenstates of $Z_v$ operators with eigenvalue 1, and similarly for $| \bfx_e = 1 \rangle$. (The boldface indicates collective notation.) The TC ground state $| \Psi_0 \rangle$ can be obtained from $| \Psi_c \rangle$ by performing forced measurements on vertex qubits~\cite{raussendorf2005long}:
\begin{align} \label{TCGS}
    | \Psi_0 \rangle \propto \langle \bfz_v = 1 | \Psi_c \rangle \propto \sum_{\bfz_e, \bfx_v} \prod_e (1 + z_e x_v x_{v'}) | \bfz_e \rangle.
\end{align}

\subsection{Doubled Hilbert Space Formalism} \label{Sec:DHSF}

Here, we review the doubled Hilbert space formalism~\cite{lee2025symmetry,bao2023mixed,lee2023quantum}, which views mixed states through the lens of pure states. For a density matrix $\rho$ defined on the Hilbert space $\HH$, the Choi-Jamio\l kowski (CJ) isomorphism maps $\rho$ to its Choi state $| \rho \rrangle$, which is a pure state on the doubled Hilbert space $\HH \otimes \bar{\HH}$, where $\bar{\HH}$ is a copy of $\HH$~\cite{jamiolkowski1972linear, choi1975completely}. More precisely, the (unnormalized) Choi state of $\rho = \sum_{i,j} \rho_{i,j} | i \rangle \langle j |$ is given by $| \rho \rrangle = \sum_{i,j} \rho_{i,j} |i, \bar{j} \rrangle$, where $|i, \bar{j} \rrangle \equiv | i \rangle | \bar{j} \rangle$ and $\{ | i \rangle \}$ is an orthonormal basis of $\HH$. (The bar notation is for the copied Hilbert space.) A quantum channel $\EE[\rho] = \sum_a K_a \rho K_a^\dagger$ with Kraus operators $\{K_a\}$ also maps under the CJ isomorphism into an operator $\bbE = \sum_a K_a \otimes \bar{K}_a^*$ acting on $\HH \otimes \bar{\HH}$.

For a pure TC state $\rho_0 = | \Psi_0 \rangle \langle \Psi_0 |$, its Choi state is given by [see Eq.~\eqref{TCGS}]
\begin{align} \label{TCGS_Choi}
    | \rho_0 \rrangle \propto \sum_{\substack{\bfz_e, \bar{\bfz}_e, \\ \bfx_v, \bar{\bfx}_v}} \prod_e (1 + z_e x_v x_{v'}) (1 + \bar{z}_e \bar{x}_v \bar{x}_{v'}) | \bfz_e, \bar{\bfz}_e \rrangle,
\end{align}
which possesses double topological order, i.e., its topological quantum field theory is the product of two $\Z_2$ TC theories. Under local decoherence described by a channel $\EE = \prod_e \EE_e$, the density state becomes $\rho_D = \EE [\rho_0]$ and its Choi state $| \rho_D \rrangle = \bbE | \rho_0 \rrangle$ may lose the double topological order depending on the decoherence strength~\cite{bao2023mixed}. From $\Tr [\rho_D^2] = \llangle \rho_D | \rho_D \rrangle = \llangle \rho_0 | \bbE^\dagger \bbE | \rho_0 \rrangle$, the purity of the decohered TC can be written as~\cite{chen2024unconventional}
\begin{align} \label{purity}
    \Tr [\rho_D^2] \propto \sum_{\bfx_v, \bar{\bfx}_v, \bft_v, \bar{\bft}_v} \prod_e \omega_e,
\end{align}
where
\begin{align} \label{weight}
    &\omega_e \equiv \sum_{z_e, \bar{z}_e, z'_e, \bar{z}'_e = \pm 1} \llangle z'_e, \bar{z}'_e | \bbE_e^\dagger \bbE_e | z_e, \bar{z}_e \rrangle \\
    &\hspace{5pt} \times (1 + z_e x_v x_{v'}) (1 + \bar{z}_e \bar{x}_v \bar{x}_{v'}) (1 + z'_e t_v t_{v'}) (1 + \bar{z}'_e \bar{t}_v \bar{t}_{v'}). \nonumber
\end{align}
The purity Eq.~\eqref{purity} can be interpreted as a partition function of a stat-mech model of Ising spins ($x_v$, $\bar{x}_v$, $t_v$, $\bar{t}_v$) on the vertices of the square lattice, with Eq.~\eqref{weight} being the local Boltzmann weight on edge $e$. Therefore, we can examine the change in double topological order under decoherence by studying the effective stat-mech model Eq.~\eqref{purity}. In Sec.~\ref{Sec:RandRotNoise} and \ref{Sec:AmpDampNoise}, we compute $\omega_e$ for coherent noise and determine the mixed-state phase diagram by examining the corresponding stat-mech model. 

\subsection{Mixed-State Phases} \label{Sec:MS_Phase}

Here, we review the concept of mixed-state phases of matter. For pure states, phase equivalence is defined as a two-way connection via a finite-depth local unitary circuit (allowing for the addition or removal of ancillary degrees of freedom in the form of product states)~\cite{verstraete2005renormalization,gu2009tensor,chen2010local}. Here, ``finite-depth'' means at most $\mathrm{polylog}(L)$ depth for a system of linear size $L$. A recent extension of this framework to mixed states proposes that two mixed states, $\rho_1$ and $\rho_2$, are in the same phase if they can be connected by a pair of finite-depth local quantum channels~\cite{coser2019classification,ma2023average,sang2024mixed,sang2025stability}, i.e., there exists finite-depth quantum channels $\mathcal{C}_1$ and $\mathcal{C}_2$ such that $\mathcal{C}_1 [\rho_1] = \rho_2$ and $\mathcal{C}_2 [\rho_2] = \rho_1$ (again, ``finite-depth'' indicates polylogarithmic depth). This definition naturally generalizes the phase equivalence of pure states and closely mirrors the original definition of Ref.~\cite{coser2019classification}, which used local Lindbladian evolution for at most polylogarithmic time.

A key distinction for mixed states is that, unlike unitary circuits for pure states, quantum channels are generally not invertible, necessitating explicit construction of both $\mathcal{C}_1$ and $\mathcal{C}_2$. In the case of local decoherence, one direction is naturally provided by the decoherence channel, whereas the reverse direction can be interpreted as a decoder for quantum memories.

For pure states, the correlation length (or energy gap) remains finite within a given phase of matter. Interestingly, a mixed-state analogue of correlation length, known as the Markov length, has been recently introduced~\cite{sang2025stability}. The Markov length scale diverges at mixed-state phase transitions, and its finiteness throughout a local Lindbladian evolution ensures the existence of a corresponding local inverse Lindbladian evolution, thereby serving the same role as correlation length in pure states.

\subsection{Diagnostics}

Local quantum channels can always be purified to a finite-depth unitary circuit on an enlarged Hilbert space with ancillary qubits so that functions linear in density matrix (e.g., $\Tr[\rho_D O]$ for some observables $O$) are smooth in tuning parameters of the local channel~\cite{lee2025symmetry, lee2023quantum, fan2024diagnostics}. Consequently, detecting phase transitions in mixed states decohered under local quantum channels requires considering functions nonlinear in density matrices, such as information-theoretic quantities. This section introduces two information-theoretic diagnostics that will be used throughout the paper.

\subsubsection{Anyon Condensation/Confinement Parameters} \label{AnyonParams}

Consider a density matrix $\rho_0^{\alpha \bar{\beta}} \equiv w_{\alpha} (l) \rho_0 w_{\beta}^\dagger (l)$, where $w_{\alpha} (l)$ ($w_{\beta} (l)$) is a string operator creating a pair of $\alpha$ ($\beta$) anyons at the endpoints $i$ and $j$ of an open string $l$. Letting $ \rho_D^{\alpha \bar{\beta}} \equiv \EE [\rho_0^{\alpha \bar{\beta}}]$, the $\alpha \bar{\beta}$ anyon condensation and confinement parameters are defined as follows~\cite{bao2023mixed, haegeman2015shadows, duivenvoorden2017entanglement}:
\begin{equation} \label{AnyonParam}
    \begin{aligned}
        \llangle I \bar{I} | \alpha \bar{\beta} \rrangle &\equiv \lim_{|i - j| \rightarrow \infty} \frac{\llangle \rho_D | \rho_D^{\alpha \bar{\beta}} \rrangle}{\llangle \rho_D | \rho_D \rrangle}, \\
        \llangle \alpha \bar{\beta} | \alpha \bar{\beta} \rrangle &\equiv \lim_{|i - j| \rightarrow \infty} \frac{\llangle \rho_D^{\alpha \bar{\beta}} | \rho_D^{\alpha \bar{\beta}} \rrangle}{\llangle \rho_D | \rho_D \rrangle}.
    \end{aligned}
\end{equation}
The condensation parameter $\llangle I \bar{I} | \alpha \bar{\beta} \rrangle$ takes a nonzero finite constant value when $\alpha \bar{\beta}$ anyons are condensed, suggesting that $\rho_0^{\alpha\bar{\beta}}$ indistinguishable from the ground state $\rho_0$ under decoherence. It vanishes when $\alpha\bar{\beta}$ anyons are not condensed. Meanwhile, the confinement parameter $\llangle \alpha \bar{\beta} | \alpha \bar{\beta} \rrangle$ vanishes when $\alpha \bar{\beta}$ anyons are confined, indicating that $| \rho_D^{\alpha \bar{\beta}} \rrangle$ is not a physically normalizable state. It attains a nonzero finite value when $\alpha\bar{\beta}$ anyons are deconfined. Note that condensing one anyon confines all other anyons with nontrivial mutual statistics, following the standard anyon condensation rule~\cite{bais2009condensate, burnell2018anyon}.

On a related note, the distinguishability between density matrices $\rho$ and $\sigma$ can be measured by the quantum relative entropy $D(\rho \| \sigma) \equiv \Tr [\rho (\log \rho - \log \sigma)]$~\cite{umegaki1962conditional}, which diverges when $\rho$ and $\sigma$ are orthogonal and takes a finite value otherwise. A useful related quantity is the R\'enyi relative entropy~\cite{petz1986quasi}:
\begin{align}
    D^{(n)} (\rho \| \sigma) \equiv \frac{1}{1 - n} \log \left( \frac{\Tr [\rho \sigma^{n-1}]}{\Tr [\rho^n]} \right),
\end{align}
which recovers $D(\rho \| \sigma)$ as $n \rightarrow 1$. The anyon condensation parameters are related to the R\'enyi-2 relative entropy as $D^{(2)} (\rho_D \| \rho_D^{\alpha \bar{\beta}}) = - \log \llangle I\bar{I} | \alpha \bar{\beta} \rrangle$. Thus, when $\alpha \bar{\beta}$ anyons are condensed (not condensed), $D^{(2)} (\rho_D \| \rho_D^{\alpha, \bar{\beta}})$ becomes finite (infinite), indicating that $\rho_D$ and $\rho_D^{\alpha,\bar{\beta}}$ are indistinguishable (orthogonal).

One can alternatively define anyon parameters by reversing the order of operations~\cite{bao2023mixed}: first applying the noise channel, then the string operators, and finally taking the overlap. For example, the anyon condensation parameter can be defined as $\llangle \rho_D | (\omega_\alpha(l) \otimes \bar{\omega}_\beta^*(l)) | \rho_D \rrangle$. We expect this parameter to behave similarly to the one given in Eq.~\eqref{AnyonParam}. This expectation is based on a field-theoretical picture of decohered toric code, where a decohering channel is represented as a temporal defect near the imaginary-time slice $\tau = 0$~\cite{bao2023mixed,garratt2023measurements}. Since both the string operator $\omega_\alpha(l) \otimes \bar{\omega}_\beta^*(l)$ and the proliferated anyons due to the defect reside on the same temporal slice, the order in which the string operator and the noise channel are applied is not expected to alter the essential physics of anyon condensation.

The anyon parameters in Eq.~\eqref{AnyonParam} for the decohered TC can be represented as observables in the stat-mech model Eq.~\eqref{purity} as follows~\cite{chen2024unconventional}:
\begin{equation} \label{AnyonParamSM}
    \begin{aligned}
        \llangle I\bar{I} | e\bar{e} \rrangle &= \lim_{|i-j| \rightarrow \infty} \langle x_i \bar{x}_i x_j \bar{x}_j \rangle, \\
        \llangle e\bar{I} | e\bar{I} \rrangle &= \lim_{|i-j| \rightarrow \infty} \langle x_i t_i x_j t_j \rangle, \\
        \llangle I\bar{I} | m\bar{m} \rrangle &= \lim_{|\tilde{i}-\tilde{j}| \rightarrow \infty} \langle \mu_{\tilde{i}}^x \mu_{\tilde{i}}^{\bar{x}} \mu_{\tilde{j}}^x \mu_{\tilde{j}}^{\bar{x}} \rangle, \\
        \llangle m\bar{I} | m\bar{I} \rrangle &= \lim_{|\tilde{i}-\tilde{j}| \rightarrow \infty} \langle \mu_{\tilde{i}}^x \mu_{\tilde{i}}^t \mu_{\tilde{j}}^x \mu_{\tilde{j}}^t \rangle, 
    \end{aligned}
\end{equation}
where $i$ and $j$ ($\tilde{i}$ and $\tilde{j}$) are the endpoints of the string defined on the original (dual) lattice. Here, $\mu^y$ (with $y = x, \bar{x}, t, \bar{t}$) is the disorder parameter of $y$ spins~\cite{kadanoff1971determination}. We review the derivation of the correspondence Eq.~\eqref{AnyonParamSM} in Appendix~\ref{App:AnyonParam}. 

\subsubsection{Coherent Information}

Let $\QQ$ represent the TC and $\RR$ be the two-qubit reference system that is maximally entangled with two logical qubits of the TC. To diagnose the amount of retrievable quantum information encoded in the TC after channel $\EE$, one can consider the coherent information $I_c (\RR \rangle \QQ) \equiv S(\EE[\rho_\QQ]) - S(\EE[\rho_{\RR\QQ}])$~\cite{schumacher1996quantum, lloyd1997capacity}, where $S(\rho) = -\Tr [\rho \log \rho]$ is the von Neumann entropy. Without decoherence (i.e., $\EE = I$), the TC achieves $I_c = 2\log 2$, a condition necessary and sufficient for the existence of an exact QEC protocol. Since $I_c$ monotonically decreases under channel $\EE$, the noise strength at which $I_c$ drops from $2\log 2$ provides an upper bound for the error threshold of any QEC scheme. Thus, investigating $I_c$ can reveal the extent to which encoded quantum information can persist under noise. 

In practice, we consider the R\'enyi coherent information~\cite{fan2024diagnostics}:
\begin{equation} \label{CI}
    \begin{aligned}
        I_c^{(n)} (\RR \rangle \QQ) &\equiv S^{(n)} (\EE[\rho_\QQ]) - S^{(n)} (\EE[\rho_{\RR\QQ}]) \\
        &= \frac{1}{n - 1} \log \left( \frac{\Tr [\EE[\rho_{\RR\QQ}]^n]}{\Tr [\EE[\rho_\QQ]^n]} \right),
    \end{aligned}
\end{equation}
where $S^{(n)} (\rho) = (1-n)^{-1} \log \Tr[\rho^n]$ is the $n$th R\'enyi entropy. As $n\rightarrow 1$, $I_c^{(n)}$ recovers $I_c$. In Appendix~\ref{App:CI}, we derive the expression for $I_c^{(2)}$ in terms of the stat-mech model [see Eq.~\eqref{MapIc} and \eqref{MapIcAT}]. Although $I_c^{(2)}$ does not give a rigorous error threshold of the code, it is relatively easy to compute and in many cases tracks the tendency of $I_c$. We argue in Appendix~\ref{App:Monotonicity} that, under certain physical assumptions, the error threshold obtained from $I_c^{(2)}$ provides an upper bound for the true error threshold corresponding to $I_c$, at which the decodability transition occurs.

\section{Random Rotation Noise} \label{Sec:RandRotNoise}

In this section, we study the decohered TC under random rotation noise. First, let's clarify the terminology before introducing the model. We refer to noise where an operator $O$ is applied with probability $p$, like the bit-flip error $\rho \mapsto (1-p) \rho + p X \rho X$, as ``stochastic $O$ noise.'' By coherent noise, we mean noise that can produce a coherent superposition of different error states (with respect to usual syndrome measurements). Thus, stochastic noise is generally coherent, except for stochastic $X$, $Y$, and $Z$ noise, which are incoherent. 

A random rotation noise is represented by a quantum channel $\EE_{\mathrm{rot}} = \prod_e \EE_e$, with
\begin{align} \label{RandRotQC}
    \EE_e [\rho] \equiv \int_{\varphi_e} g(\varphi_e) U_e (\varphi_e) \rho U_e^\dagger (\varphi_e),
\end{align}
where $U_e (\varphi_e) = e^{-i\varphi_e (\vec{n} \cdot \vec{\sigma})_e}$ and $\int_{\varphi_e} \equiv \int_{-\pi}^\pi d\varphi_e$. Here, each qubit is independently rotated around the axis $\vec{n} = (n_x, n_y, n_z)$ by an angle $\varphi_e$ drawn from the probability distribution $g(\varphi_e)$ defined on $[-\pi, \pi)$ [see Fig.~\ref{fig:Setup_Summary}(a)]. Such noise often arises in current quantum processors due to imperfect gate operations~\cite{kueng2016comparing,greenbaum2017modeling,bravyi2018correcting,venn2023coherent,tomita2014low,hakkaku2021sampling,pataki2024coherent,marton2023coherent,venn2020error}. The case with $\vec{n} = \hat{z}$ and an even distribution $g(\varphi_e)$ has been studied in Ref.~\cite{pataki2024coherent} under the name ``quasistatic phase damping noise.'' In this work, we are interested in $\rho_D = \EE_{\mathrm{rot}} [\rho_0]$ and we impose no constraints on the rotation axis $\vec{n}$ or the distribution $g(\varphi_e)$.

\begin{figure}[t]
    \includegraphics[width=\columnwidth]{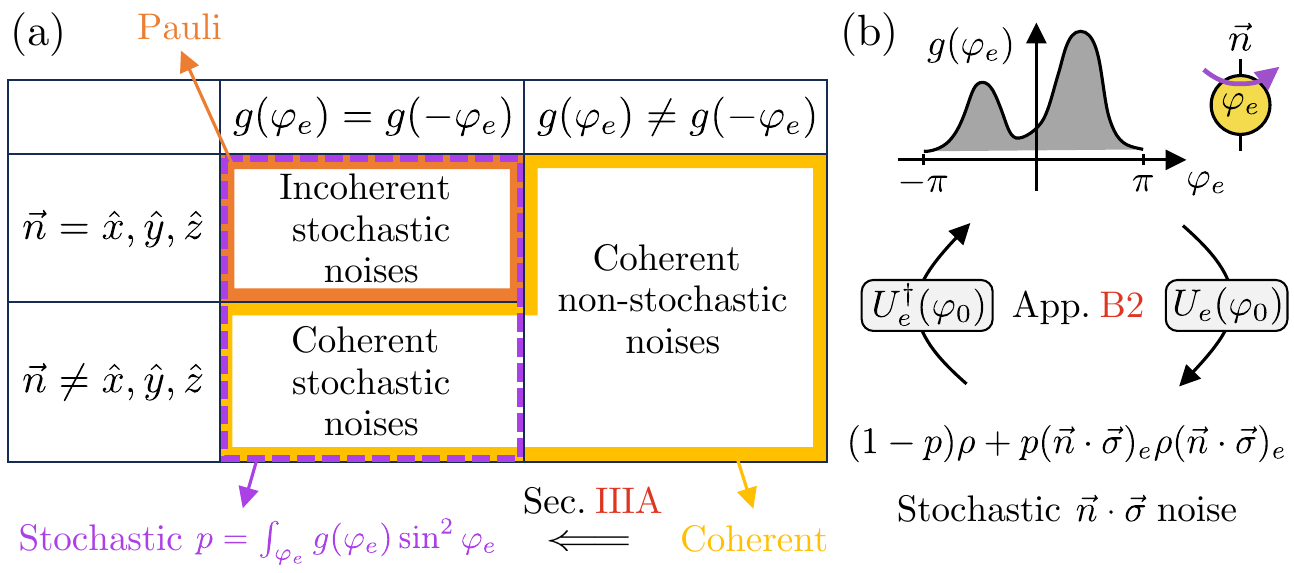}
    \caption{(a) Summary of the properties of the random rotation noise Eq.~\eqref{RandRotQC}, based on its angle distribution $g(\varphi_e)$ and rotation axis $\vec{n}$. Sec.~\ref{Sec:MapSMModel} shows that it is sufficient to study stochastic $\vec{n} \cdot \vec{\sigma}$ noise to map out mixed-state phase diagram for any distribution $g(\varphi_e)$. (b) Random rotation noise with any distribution can be transformed into stochastic $\vec{n}\cdot \vec{\sigma}$ noise via a depth-1 unitary $U_e(\varphi)$, with $\varphi_0$ given in Eq.~\eqref{varphi0} (see Appendix~\ref{App:RDependence} for detail). Consequently, both distributions yield the same mixed-state phase.}
    \label{fig:Noise_Distn_Relation}
\end{figure}

In general, the random rotation noise Eq.~\eqref{RandRotQC} is coherent and not stochastic. However, when the distribution is even [i.e., $g(\varphi_e) = g(-\varphi_e)$], Eq.~\eqref{RandRotQC} reduces to stochastic $\vec{n} \cdot \vec{\sigma}$ noise with an error rate $p = \int_{\varphi_e} g(\varphi_e) \sin^2 \varphi_e$ (see Appendix~\ref{App:ReducStoNoise} for a proof). Thus, Eq.~\eqref{RandRotQC} encompasses previously studied incoherent Pauli noise, such as bit-flip and phase-flip noise. See Fig.~\ref{fig:Noise_Distn_Relation}(a) for a summary of the properties of random rotation noise according to its angle distribution $g(\varphi_e)$ and rotation axis $\vec{n}$.

A global uniform rotation of qubits by an angle $\epsilon$ corresponds to a Dirac-delta distribution $g(\varphi_e) = \delta(\varphi_e - \epsilon)$. However, given the ability to learn \emph{a priori} the angle distribution during qubit calibration, this error can be corrected by rotating back each qubit by $-\epsilon$. In practice, 
however, $g(\varphi_e)$ may be dispersed, making simple counter-rotations ineffective. In this sense, random rotation noise is a practically motivated noise model, and understanding error thresholds for such noise with general distributions is crucial---especially when counteracting unitaries can be applied to the decohered state before syndrome measurements. This possibility is naturally incorporated in the notion of the mixed-state phase, which remains invariant under a finite-depth local channel (see Sec.~\ref{Sec:MS_Phase}).

\subsection{Mapping to Statistical Mechanics Model} \label{Sec:MapSMModel}

One may wonder how to analyze the wide variety of angle distributions. We find that a remarkable simplification arises by allowing the application of single-qubit unitaries. More precisely, we show in Appendix~\ref{App:RDependence} that random rotation noise with rotation axis $\vec{n}$ and an arbitrary distribution $g(\varphi_e)$ can be always transformed via single-qubit unitary rotations into stochastic $\vec{n} \cdot \vec{\sigma}$ noise, which may be of independent interest in QEC applications. The error rate of the resulting stochastic $\vec{n} \cdot \vec{\sigma}$ noise is given by $p = (1 - \sqrt{1 - R}) / 2$, with
\begin{align} \label{R}
    R \equiv 1 - \left| \int_{\varphi_e} g(\varphi_e) e^{2i\varphi_e} \right|^2.
\end{align}
The parameter $R$ quantifies the degree of decoherence induced by the distribution. For example, we have $R = 0$ for any global trivial rotation with $g(\varphi_e) = \delta(\varphi_e - \epsilon)$, and $R = 1$ for maximally random noise with $g(\varphi_e) = (2\pi)^{-1}$.

Since the mixed state phase does not change under such local unitary, we can study the mixed-state phase of the decohered TC under a general angle distribution $g(\varphi_e$) by focusing solely on the stochastic $\vec{n} \cdot \vec{\sigma}$ noise (which corresponds to even distributions). Namely, all the information about $g(\varphi_e)$ is encoded in the single parameter $R$; distributions with the same value of $R$ belong to the same mixed-state phase. The phase diagram obtained from this noise can be directly translated for general distributions by expressing $R$ in terms of parameters characterizing the distribution (see Sec.~\ref{Sec:ExDvM}).

As discussed in Sec.~\ref{Sec:DHSF}, the mixed-state topological order within the doubled Hilbert space formalism can be investigated by expressing the purity $\Tr[\rho_D^2] = \llangle \rho_0 | \bbE_{\mathrm{rot}}^\dagger \bbE_{\mathrm{rot}} | \rho_0 \rrangle$---which also depends on the distribution $g(\varphi_e)$ solely through $R$---as a partition function of an effective stat-mech model. (This strategy applies to all $n$th moments of the density matrix, $\Tr [\rho_D^n]$, including the replica-limit case $n = 1$.) Under the CJ isomorphism, the channel Eq.~\eqref{RandRotQC} is mapped to $\bbE_{\mathrm{rot}} = \prod_e \bbE_e$, where
\begin{align} \label{RandRotCJ}
    \bbE_e = \int_{\varphi_e} g(\varphi_e) e^{-i\varphi_e (\vec{n} \cdot \vec{\sigma})_e} \otimes \overline{e^{i\varphi_e (\vec{n}' \cdot \vec{\sigma})_e}} 
\end{align}
and $\vec{n}' = (n_x, -n_y, n_z)$ is introduced for the copied Hilbert space. Without loss of generality, we assume an even distribution $g(\varphi_e)$. We show in Appendix~\ref{App:StatMechMap} that the purity $\Tr[\rho_D^2]$ in the thermodynamic limit ($L\rightarrow \infty$) is proportional to the partition function $\sum_{\bfs_v, \bftau_v} \prod_e \omega_e$ with
\begin{align} \label{SMModel}
    \omega_e \propto 1 + J_1 s_v s_{v'} + J_2 \tau_v \tau_{v'} + K s_v s_{v'} \tau_v \tau_{v'},
\end{align}
where $s_v \equiv x_v \bar{x}_v$ and $\tau_v \equiv \bar{x}_v t_v$ are the Ising spins residing on vertices of a square lattice, and
\begin{equation} \label{weightRandRot}
    \begin{aligned}
        J_1 &= \frac{R (\sin^2 \theta \cos^2 \phi + \cos^2 \theta)}{2 - R + R \sin^2 \theta \sin^2 \phi}, \\
        J_2 &= \frac{R (\sin^2 \theta \cos^2 \phi - \cos^2 \theta)}{2 - R + R \sin^2 \theta \sin^2 \phi}, \\
        K &= \frac{2 - R - R \sin^2 \theta \sin^2 \phi}{2 - R + R \sin^2 \theta \sin^2 \phi}.
        \end{aligned}
\end{equation}
Here, we parameterize the rotation axis as $\vec{n} = (\sin\theta \sin\phi, \cos\theta, \sin\theta \cos\phi)$, which is the spherical coordinates with the $Y$-axis as the polar axis [see Fig.~\ref{fig:Setup_Summary}(b)]. We can rewrite Eq.~\eqref{SMModel} as a local Boltzmann weight $\omega_e \propto e^{\mathcal{J}_1  s_v s_{v'} + \mathcal{J}_2 \tau_v \tau_{v'} + \mathcal{K} s_v s_{v'} \tau_v \tau_{v'}}$. The three coupling constants $\JJ_1$, $\JJ_2$, $\KK$ are determined by $J_1$, $J_2$, $K$, and can generally be anisotropic and complex (see Appendix~\ref{App:StatMechMap} for details). In particular, the model gains its non-Hermiticity as the $y$-component of $\vec{n}$ increases. Consequently, $\Tr[\rho_D^2]$ is mapped to the partition function of the non-Hermitian anisotropic Ashkin-Teller (AT) model. This is one of our main results and we emphasize that this mapping can be leveraged to investigate decohered TC under random rotation noise with \emph{any} rotation axis $\vec{n}$ and \emph{arbitrary} angle distribution $g(\varphi_e)$.

From Eq.~\eqref{AnyonParamSM}, the anyon parameters can be read as
\begin{equation} \label{AP_RandRot}
    \begin{aligned}
        \llangle I\bar{I} | e\bar{e} \rrangle &= \lim_{|i-j| \rightarrow \infty} \langle s_i s_j \rangle, \\
        \llangle e\bar{I} | e\bar{I} \rrangle &= \lim_{|i-j| \rightarrow \infty} \langle s_i \tau_i s_j \tau_j \rangle, \\
        \llangle I\bar{I} | m\bar{m} \rrangle &= \lim_{|\tilde{i}-\tilde{j}| \rightarrow \infty} \langle \mu_{\tilde{i}}^\tau \mu_{\tilde{j}}^\tau \rangle, \\
        \llangle m\bar{I} | m\bar{I} \rrangle &= \lim_{|\tilde{i}-\tilde{j}| \rightarrow \infty} \langle \mu_{\tilde{i}}^s \mu_{\tilde{i}}^\tau \mu_{\tilde{j}}^s \mu_{\tilde{j}}^\tau \rangle,
    \end{aligned}
\end{equation}
which are the correlation functions in the model Eq.~\eqref{SMModel}. The stat-mech expression for the R\'enyi-2 coherent information is given by 
\begin{align} \label{RandRotIc}
    I_c^{(2)} (\RR\rangle \QQ) &= \log \left( \frac{\sum_{a,b = 0,1} e^{-\Delta F_{s,\tau}^{(a,b)}}}{\sum_{a,b = 0,1} e^{-\Delta F_{\tau,s\tau}^{(a,b)}}} \right),
\end{align}
where $\Delta F_{s,\tau}^{(a,b)}$ is a free energy cost of forming defects in the interactions $s_v s_{v'}$ and $\tau_v \tau_{v'}$ along the non-contractible loop pattern $(a,b)$, and similarly for $\Delta F_{\tau, s\tau}^{(a,b)}$ (see Appendix.~\ref{App:CI} for details). Using these two diagnostics, we can determine the phase of the model Eq.~\eqref{SMModel} (and hence that of $\rho_D$) and determine the fate of the quantum memory under random rotation noise.

\subsection{General Phase Diagram} \label{Sec:GeneralPD}

In the following, we numerically map out the general phase diagram of the stat-mech model Eq.~\eqref{SMModel} with respect to $(R, \phi, \theta)$. From the symmetry of Eq.~\eqref{weightRandRot}, it suffices to study the parameter regime $0\leq R\leq 1$ and $0 \leq \phi,\theta \leq \pi/2$. We represent the partition function of the stat-mech model into a tensor network and employ the corner transfer matrix renormalization group (CTMRG) algorithm~\cite{nishino1995density, nishino1996corner, fishman2018faster} to approximately contract the tensor network. The phase boundaries are determined by pinpointing the parameters where the correlation length diverges and the order parameters change (see Fig.~\ref{fig:CTMRG_RandRot_OPCorr}). The result is summarized in Fig.~\ref{fig:Setup_Summary}(b), where the phase boundaries are marked by yellow and green surfaces. We have confirmed that the correlation length $\xi_D$ increases as we increase the bond dimension $D$ of the corner tensors (not shown). The model has the following three phases:

\begin{figure}[t]
    \includegraphics[width=\columnwidth]{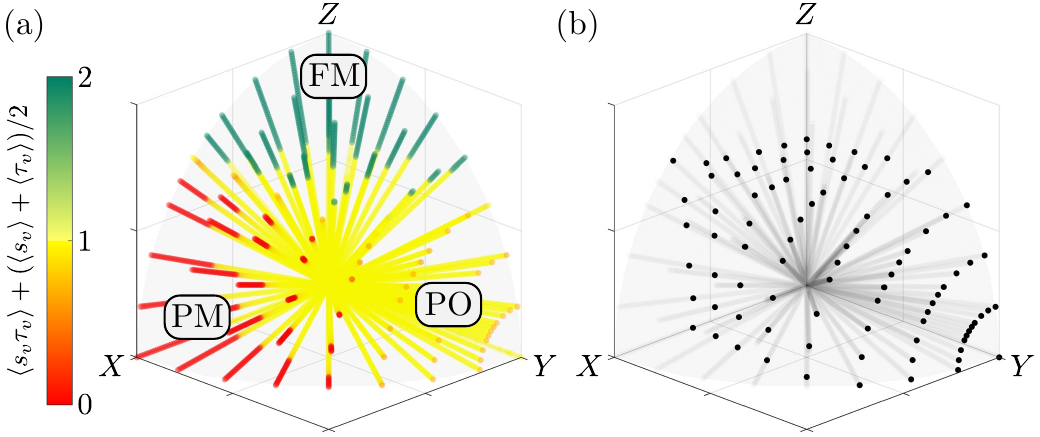}
    \caption{CTMRG results for the statistical mechanics model Eq.~\eqref{SMModel}. (a) shows $\langle s_v \tau_v \rangle + ( \langle s_v \rangle + \langle \tau_v \rangle) / 2$, which takes the values close to 0, 1, and 2 for the paramagnetic (PM), partially ordered (PO), and ferromagnetic (FM) phases, respectively. (b) shows the points where the correlation length achieves the maximum value for each ray (gray lines). The bond dimension used is $D = 40$. Here, the radial distance represents the parameter $R$, ranging from 0 (at the origin) to 1.}
    \label{fig:CTMRG_RandRot_OPCorr}
\end{figure}

\begin{enumerate}[leftmargin=*]
    \item Partially ordered (PO) phase = Quantum memory: \\
    In this case, $\langle s_v \tau_v \rangle \neq 0$ and $\langle s_v \rangle = \langle \tau_v \rangle = 0$ so that $\llangle I\bar{I} | e\bar{e} \rrangle = \llangle I\bar{I} | m\bar{m} \rrangle = 0$ and $\llangle e\bar{I} | e\bar{I} \rrangle$, $\llangle m\bar{I} | m\bar{I} \rrangle \neq 0$. Thus, the PO phase corresponds to double $\Z_2$ topological order and $\rho_D$ hosts quantum memory. Indeed, the pure TC with $R = 0$ yields $\JJ_1 = \JJ_2 = 0$ and $\KK = \infty$, which belongs to the PO phase.

    \item Ferromagnetic (FM) phase = Classical Memory: \\
    In this case, $\langle s_v \tau_v \rangle$, $\langle s_v \rangle$, $\langle \tau_v \rangle \neq 0$ so that $\llangle I\bar{I} | m\bar{m} \rrangle = \llangle m\bar{I} | m\bar{I} \rrangle = 0$ and $\llangle I\bar{I} | e\bar{e} \rrangle$, $\llangle e\bar{I} | e\bar{I} \rrangle \neq 0$, indicating that $e\bar{e}$ anyons are condensed and $m$ anyons are confined. Thus, the FM phase corresponds to a single $\Z_2$ topological order with the $e\bar{e}$ anyon condensed, leaving classical memory in $\rho_D$.

    \item Paramagnetic (PM) phase = Classical memory: \\
    In this case, $\langle s_v \tau_v \rangle = \langle s_v \rangle = \langle \tau_v \rangle = 0$ so that $\llangle I\bar{I} | e\bar{e} \rrangle = \llangle e\bar{I} | e\bar{I} \rrangle = 0$ and $\llangle I\bar{I} | m\bar{m} \rrangle$, $\llangle m\bar{I} | m\bar{I} \rrangle \neq 0$. This indicates that $m\bar{m}$ anyons are condensed and $e$ anyons are confined. Thus, the PM phase corresponds to a single $\Z_2$ topological order with the $m\bar{m}$ anyon condensed, leaving classical memory in $\rho_D$.
\end{enumerate}

One can readily check that the behavior of $I_c^{(2)}$ in Eq.~\eqref{RandRotIc} is consistent with the above analysis based on the anyon parameters. For example, consider the PO phase where $s$ and $\tau$ spins are disordered so that $\Delta F_{s,\tau}^{(a,b)} = 0$. Also, since $\langle s_v \tau_v \rangle \neq 0$, $\Delta F_{\tau,s\tau}^{(a,b)} = \OO(L)$ unless $a = b = 0$, in which case $\Delta F_{\tau,s\tau}^{(0,0)} = 0$. Therefore, $I_c^{(2)} = 2\log 2$ and hence QEC is possible in the PO phase, in agreement with the presence of quantum memory. Similarly, one can see that $I_c^{(2)} = 0$ in both the FM and PM phases, signifying a loss of quantum memory.

The region of double $\Z_2$ topological order is a bulk enclosed by the yellow and green surfaces in Fig.~\ref{fig:Setup_Summary}(b), which corresponds to the PO phase and contains the pure TC at the origin (blue dot). The two disjoint regions beyond the yellow phase boundaries correspond to single $\Z_2$ topological order with $e\bar{e}$ and $m\bar{m}$ anyons condensed, respectively. These phases correspond to the FM and the PM phases. 

\begin{figure}[t]
    \includegraphics[width=\columnwidth]{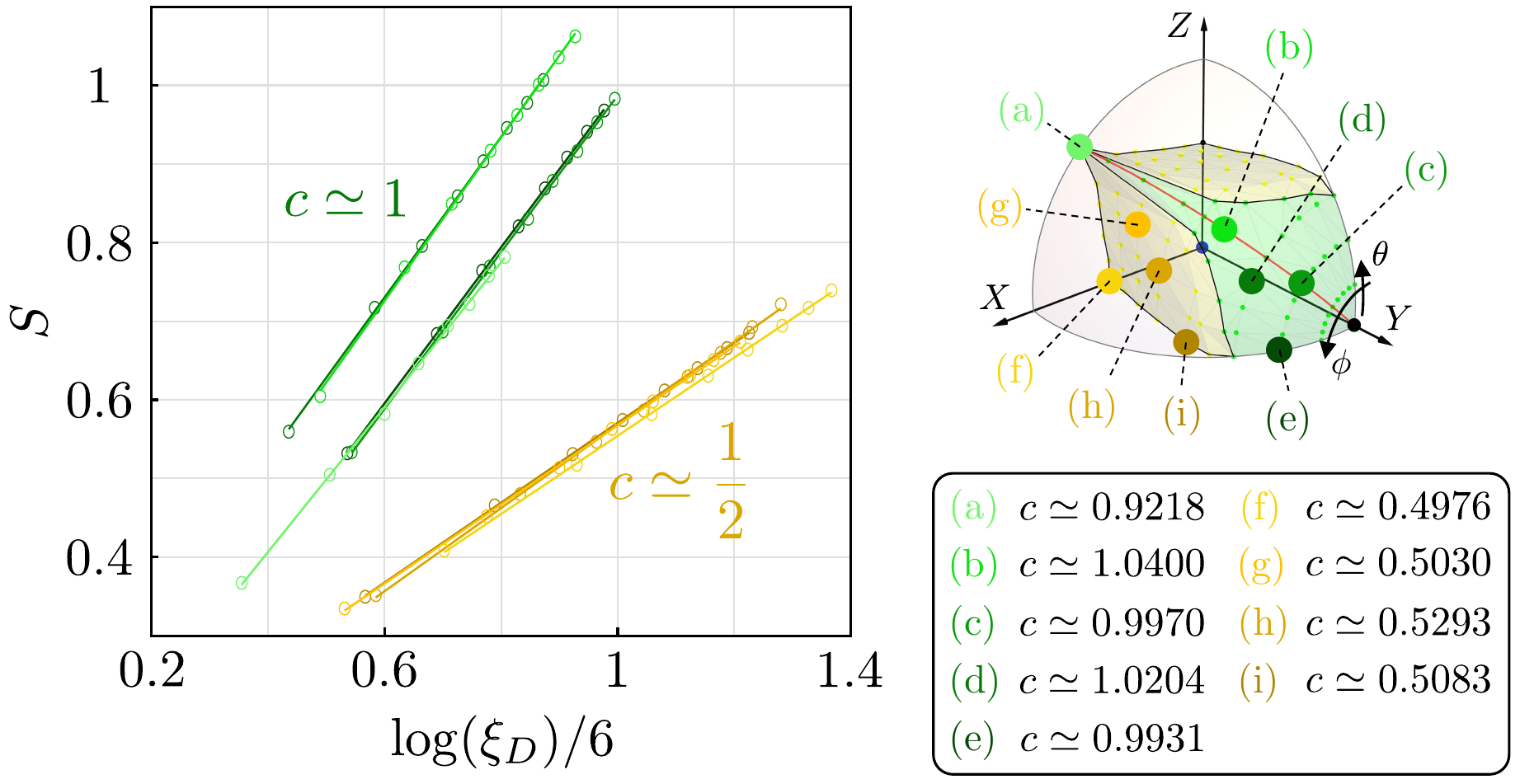}
    \caption{Finite-entanglement scaling $S \sim (c/6) \log \xi_D$ for the model Eq.~\eqref{SMModel}. Here, $\xi_D$ is a correlation length computed via CTMRG with bond dimension $10 \leq D \leq 80$. The plotted critical points correspond to the rotation axes $\vec{n}$ with (a) $\phi = \pi/4, \theta = \pi/2$, (b) $\phi = \theta = \pi/4$, (c) $\phi = \pi/4, \theta = \pi/8$, (d) $\phi = 5\pi/16, \theta = 3\pi/16$, (e) $\phi = \pi/2, \theta = \pi/8$, (f) $\phi = \theta = \pi/2$, (g) $\phi = 5\pi/16, \theta = 3\pi/8$, (h) $\phi = 3\pi/8, \theta = 5\pi/16$, and (i) $\phi = \pi/2, \theta = \pi/4$.}
    \label{fig:CTMRG_RandRot_c}
\end{figure}

To determine the universality of the phase boundaries, we compute the half-space entanglement entropy $S$ of the ground state whose Hamiltonian is determined by the transfer matrix of the model Eq.~\eqref{SMModel}. At the critical point, the tensor network with finite bond dimension $D$ cannot represent the diverging entanglement. However, finite $D$ leads to a systematic error in the captured entanglement so that $S \sim (c/6) \log \xi_D$ holds, where $c$ is a central charge of the underlying conformal field theory~\cite{pollmann2009theory}. Using this finite-entanglement scaling, we extract the central charges $c$ at the phase boundaries as shown in Fig.~\ref{fig:CTMRG_RandRot_c}. The whole yellow boundaries belong to the 2D Ising criticality with $c = 1/2$. On the other hand, the green boundary at $R = 1$ surrounding the $\phi = \pi/4$ curve (except for the black dot representing $\hat{n} = \hat{y}$) is critical with the extracted central charge close to $c = 1$. 

Below, we give several remarks regarding the cases where Eq.~\eqref{SMModel} is analytically treatable, connections to previous studies, the critical regions with $c = 1/2$ and $c = 1$, and the phase diagram in the replica limit. 

\subsubsection{Random Rotation Noise within $XZ$-Plane} \label{Sec:RandRotXZ}

When $\vec{n} = (\sin\phi, 0, \cos\phi)$ lies in the $XZ$-plane, the model Eq.~\eqref{SMModel} reduces to $\omega_e \propto 1 + J (s_v s_{v'} + \tau_v \tau_{v'}) + K s_v s_{v'} \tau_v \tau_{v'}$ with
\begin{align}
    J = \frac{R \cos^2 \phi}{2 - R \cos^2 \phi}, \qquad K = \frac{2 - R - R \sin^2 \phi}{2 - R \cos^2 \phi},
\end{align}
which can be written as the Boltzmann weight $\omega_e \propto e^{\mathcal{J} (s_v s_{v'} + \tau_v \tau_{v'}) + \mathcal{K} s_v s_{v'} \tau_v \tau_{v'}}$ of the isotropic AT model with coupling constants $\JJ, \KK \geq 0$. It is known that this model has three distinct phases (PO, FM, PM)~\cite{baxter2007exactly}, which is in accord with the $XZ$-plane of Fig.~\ref{fig:Setup_Summary}(b) and Fig.~\ref{fig:CTMRG_RandRot_OPCorr}. For the pure $X$- and $Z$-rotations, the model simplifies further to a 2D Ising model on a square lattice, with a phase transition at $J_c = \frac 12 \ln(1 + \sqrt{2})$~\cite{onsager1944crystal, kramers1941statistics}. Thus, the phase boundary for the $X$- and $Z$-rotations is $R_c = 2 - \sqrt{2} \simeq 0.586$, which belongs to 2D Ising universality and extends into two Ising critical lines for $\phi \neq \pi/4$~\cite{baxter2007exactly}.

The special cases with $(\phi, \theta) = (\pi/4, \pi/2)$ satisfy the self-duality condition $e^{-2\KK} = \sinh (2\JJ)$~\cite{baxter2007exactly}. In this scenario, the model remains in the PO phase for $R < 1$ and reaches the Berezinskii-Kosterlitz-Thouless (BKT) transition point at $R = 1$ [green dot in Fig.~\ref{fig:Setup_Summary}(b)]. This BKT transition was first discussed in Ref.~\cite{chen2024unconventional} for the stochastic channel $\EE_e [\rho] = (1-p) \rho + p (\vec{n} \cdot \vec{\sigma})_e \rho (\vec{n} \cdot \vec{\sigma})_e$ (for which $R = 1$ corresponds to $p = 1/2$), and our result greatly extends such transition for general distributions $g(\varphi_e)$. It is shown in Ref.~\cite{chen2024unconventional} that any transition in $| \rho_D \rrangle$ is beyond the conventional anyon condensation scheme if $\EE$ satisfies (i) electromagnetic duality (EMD) symmetry: $\mathbb{U}_D \bbE \mathbb{U}_D^\dagger = \bbE$ for $\mathbb{U}_D = \prod_e \frac 12 (X_e + Z_e) (\bar{X}_e + \bar{Z}_e)$, and (ii) partial transpose symmetry: $(\bbE^\dagger \bbE)^{T_\HH} = (\bbE^\dagger \bbE)^{\bar{T}_{\bar{\HH}}} = \bbE^\dagger \bbE$. One can easily check that $\bbE_{\mathrm{rot}}$ [see Eq.~\eqref{RandRotCJ}] with $(\phi, \theta) = (\pi/4, \pi/2)$ satisfies both the EMD and partial transpose symmetries. Therefore, this BKT transition is an unconventional topological phase transition for any $g(\varphi_e)$ achieving $R = 1$. Such a case involves a non-stochastic noise $\EE_{\mathrm{rot}}$ with respect to asymmetric distributions with vanishing second Fourier coefficient, e.g., $g(\varphi_e) = (2\pi)^{-1} (1 + \sin \varphi_e \cos 2\varphi_e)$.

\subsubsection{Random $Y$-Rotation Noise} \label{Sec:RandYRot}

The green phase boundaries with $R = 1$ in Fig.~\ref{fig:Setup_Summary}(b) demonstrate that the mixed-state topological order is remarkably stable against random rotation noise with $\vec{n}$ near the $Y$-axis ($\theta = \phi = 0$), suggesting that the QEC is possible in principle for the TC under such noise. This stability can be understood analytically for the pure $Y$-rotation, for which the model Eqs.~\eqref{SMModel} becomes 
\begin{align} \label{Weight_Y}
    \omega_e \propto 1 + \frac{R}{2 - R} (s_v s_{v'} - \tau_v \tau_{v'}) + s_v s_{v'} \tau_v \tau_{v'}.
\end{align}
In Appendix~\ref{App:StagVertexModel}, we map Eq.~\eqref{Weight_Y} to a certain vertex model called the ``staggered vertex'' model. We analytically compute the transfer matrix and examine the correlation length, order parameter, and anyon parameters (see Appendix.~\ref{App:StagVertexModel} for details). We find that the model has no phase transition for $R < 1$ and undergoes spontaneous symmetry breaking at $R = 1$. This suggests that double $\Z_2$ topological order persists for $R < 1$, which is consistent with our numerically obtained phase diagram [see Fig.~\ref{fig:Setup_Summary}(b) and Fig.~\ref{fig:CTMRG_RandRot_OPCorr}].

\begin{figure}[t]
    \centering
    \includegraphics[width=0.8\columnwidth]{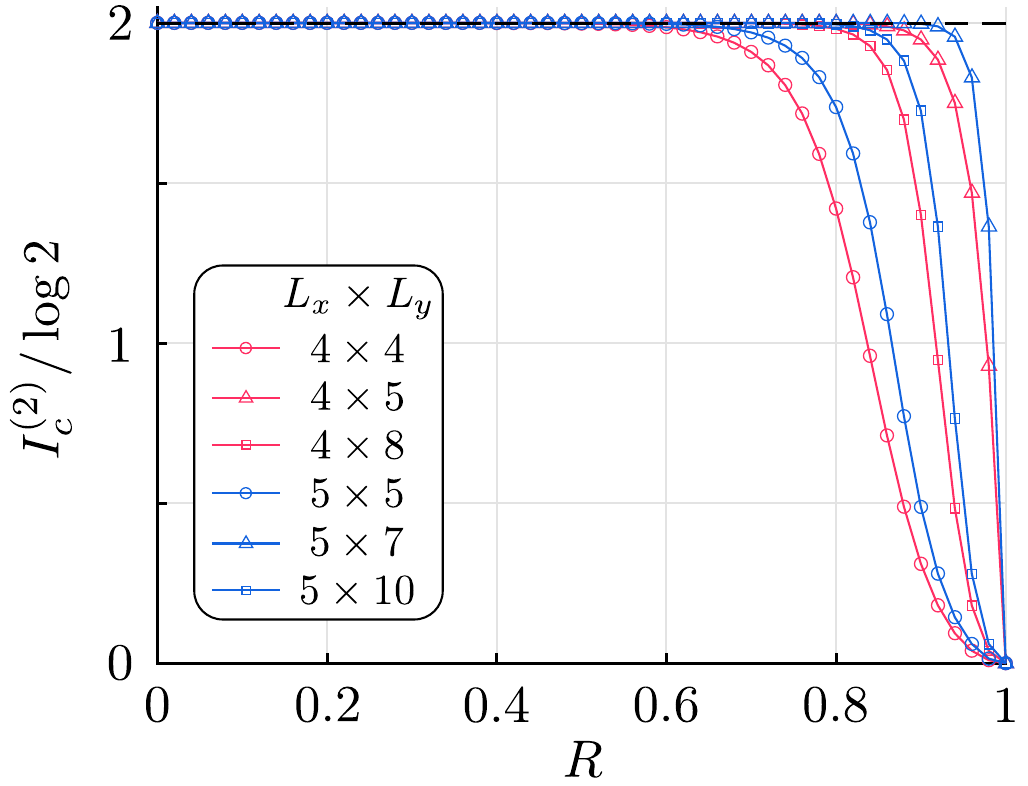}
    \caption{R\'enyi-2 coherent information $I_c^{(2)}$ of decohered toric code under random $Y$-rotation noise on various $L_x \times L_y$ torus geometry.}
    \label{fig:YRot_Renyi2CI}
\end{figure}

For coherent information, we numerically compute $I_c^{(2)}$ from Eq.~\eqref{RandRotIc} for finite-sized systems. Figure~\ref{fig:YRot_Renyi2CI} shows the result for various $L_x \times L_y$ torus geometries. Even in small systems, $I_c^{(2)}$ converges quite well to $2\log 2$ for $R < 1$. At $R = 1$, we have $I_c^{(2)} = 0$ and the quantum memory is lost. An intriguing point to note is that $I_c^{(2)}$ goes to 1 rapidly for geometries with $\gcd(L_x, L_y) = 1$. This aligns with the fact that the TC with coprime geometry has an advantage in correcting stochastic Pauli-$Y$ noise~\cite{tuckett2019tailoring}.

Although $I_c^{(2)}$ does not rigorously guarantee the existence of the QEC protocol, we expect the same behavior to persist in the replica limit $n\rightarrow 1$. This anticipation is consistent with previous studies that demonstrated the error threshold of the TC against stochastic $Y$ noise to be $p_c = 0.5$~\cite{tuckett2018ultrahigh, tuckett2019tailoring, chubb2021statistical}. Additionally, the decohered TC under the maximal stochastic $Y$ noise has a diverging Markov length~\cite{ellison2025toward}, which has been proposed to play a role similar to the correlation length in pure states~\cite{sang2025stability}. This fact suggests that there should be a mixed-state phase transition at $R_c = 1$. 

We also note similar robustness of the TC under the only-$Y$ measurement setup, where each qubit is measured in the $Y$-basis with probability $p_y$. It is known that the probability of corrupting the encoded quantum memory vanishes exponentially with system size unless every physical qubit is measured (i.e., $p_y = 1$)~\cite{botzung2025robustness, lee2024randomly}, in agreement with the mixed-state phase transition of the TC under random $Y$-rotation noise at $R_c = 1$.

\subsubsection{Critical Region with \texorpdfstring{$c = 1/2$}{c = 1/2}} \label{Sec:CriticalYellow}

The Ising critical lines identified within the $XZ$ plane in Sec.~\ref{Sec:RandRotXZ} are part of the yellow phase boundaries in Fig.~\ref{fig:Setup_Summary}(b), which separate regions of double and single $\Z_2$ topological orders. Notably, these entire two-dimensional phase boundaries exhibit Ising criticality ($c = 1/2$). We believe that this arises from the non-Hermiticity of the effective stat-mech model Eq.~\eqref{SMModel}, which at the field-theory level may introduce an imaginary mass term that is reported to extend the Ising criticality~\cite{krvcmar2022ising}. This instance illustrates the importance of studying non-Hermitian stat-mech models to understand quantum many-body states under decoherence, a direction that deserves further investigation.

\subsubsection{Critical Region with \texorpdfstring{$c = 1$}{c = 1}} \label{Sec:CriticalGreen}

Spreading out from the BKT transition point at $\vec{n}_{\text{BKT}} = (1/\sqrt{2}, 0, 1/\sqrt{2})$ [green dot in Fig.~\ref{fig:Setup_Summary}(b)], the green phase boundaries with $R = 1$ form an extended critical region with a central charge $c = 1$, except for the pure $Y$-rotation case $\vec{n}_Y = (0, 1, 0)$. We expect that this extended critical region may be described by a non-unitary conformal field theory with an effective central charge $c_* = 1$. It will be interesting to further investigate the precise nature of this critical region for future work.

We expect that the mixed-state transitions at the green phase boundary cannot be described by the standard anyon condensation scheme, as in the BKT transition point $\vec{n}_{\text{BKT}}$. A notable point is that Eq.~\eqref{RandRotCJ} with a general rotation axis $\vec{n}$ pointing within the green critical region does not satisfy the EMD and/or partial-transpose symmetries discussed in Sec.~\ref{Sec:RandRotXZ}. This suggests that this critical region may provide an example of an unconventional mixed-state phase transition beyond the sufficient conditions for such a transition derived in Ref.~\cite{chen2024unconventional}, which deserves future investigation.

We comment on a similar observation in the Hamiltonian setting, where the fixed-point toric code Hamiltonian [Eq.~\eqref{TC}] is perturbed by the term $-\sum_e (\vec{h} \cdot \vec{\sigma})_e$ with $\vec{h} = (h_x, h_y, h_z)$ representing a uniform magnetic field. In this model, a critical line with continuously varying critical exponents is found along the line $h_x = h_z$ as $h_y$ changes~\cite{dusuel2011robustness}, which suggests a connection to an underlying Ashkin-Teller-type model. Such critical line is notably analogous to the $\phi = \pi/4$ curve in our random rotation noise model [see Fig.~\ref{fig:Setup_Summary}(b)]. 

\subsubsection{Phase Diagram in the Replica Limit} \label{Sec:PDReplicaLimit}

Finally, we discuss the potential structure of the phase diagram of $\rho_D$ in the replica limit $n\rightarrow 1$. While the doubled Hilbert space formalism
offers insights into phase structures, the actual phase boundaries and universality for $n = 1$ may differ. For example, the TC under incoherent $X$/$Z$ noise follows the 2D random-bond Ising model (RBIM) universality~\cite{dennis2002topological, fan2024diagnostics, lee2025exact, niwa2025coherent} rather than the 2D Ising universality obtained from the doubled Hilbert space formalism. Based on this, we conjecture that the yellow phase boundaries fall within the 2D RBIM universality class for $n = 1$. Another interesting question is whether the green phase boundary remains critical at $n = 1$ and how its shape changes. Since the phase boundary for pure $Y$-rotation appears to stay at $R_c = 1$ in the replica limit (see Sec.~\ref{Sec:RandYRot}), one might consider three potential scenarios: (i) two isolated points at $\vec{n}_{\text{BKT}}$ and $\vec{n}_Y$ with $R = 1$, (ii) a critical curve $\phi = \pi/4$ connecting these two points, and (iii) a critical region with these points as its vertices. In each case, we believe that the pure $Y$-rotation point $\vec{n}_Y$ is not critical.

\subsection{Example: Double von Mises Distribution} \label{Sec:ExDvM}

\begin{figure}[t]
    \includegraphics[width=\columnwidth]{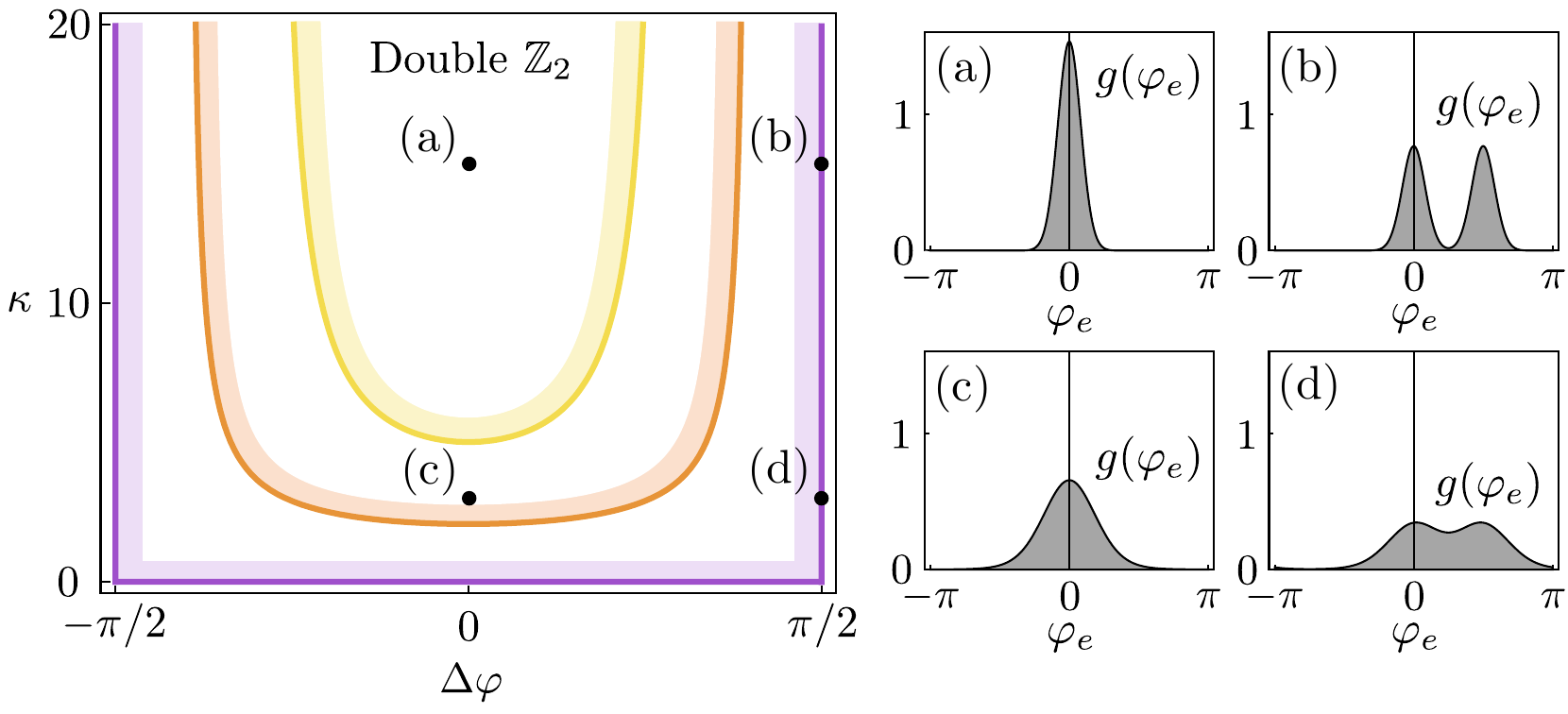}
    \caption{Phase diagram of toric code under random rotation noise with respect to the double von-Mises distribution $g_{\kappa, \Delta \varphi} (\varphi_e)$ [see Eq.~\eqref{DoublevM}]. The yellow/orange/purple curves correspond to phase boundaries for $\vec{n} = (1, 0, 0)$ / $(1 / \sqrt{2}, 1 / \sqrt{2}, 0)$ / $(0, 1, 0)$, respectively. The double (single) $\Z_2$ topological order is within the region (not) surrounded by the shade for each $\vec{n}$. The distribution profiles on the right show $g_{\kappa, \Delta \varphi} (\varphi_e)$ with (a) $\kappa = 15, \Delta \varphi = 0$, (b) $\kappa = 15, \Delta \varphi = \pi/2$, (c) $\kappa = 3, \Delta \phi = 0$, and (d) $\kappa = 3, \Delta \varphi = \pi/2$.}
    \label{fig:Ex_DoublevMDist}
\end{figure}

In this section, we provide an exemplary mixed-state phase diagram for parameters characterizing a certain angle distribution. A von Mises (vM) distribution is defined as
\begin{align} \label{vM}
    h_{\kappa, \varphi} (\varphi_e) = \frac{e^{\kappa \cos(\varphi_e  - \varphi)}}{2\pi I_0(\kappa)},
\end{align}
where $I_\alpha(\kappa)$ is the modified Bessel function of the first kind of order $\alpha$. The distribution Eq.~\eqref{vM} has a mean $\varphi$ and a variance $1 - I_1(\kappa) / I_0(\kappa)$. Since $1 - I_1(\kappa) / I_0(\kappa) \rightarrow (2\kappa)^{-1}$ as $\kappa \rightarrow \infty$, we can interpret $\kappa^{-1}$ as a measure of concentration. As $\kappa \rightarrow 0$ ($\infty$), Eq.~\eqref{vM} approaches the uniform (Dirac delta) distribution on $[-\pi, \pi)$. We consider a double vM distribution $g(\varphi_e)$, a convex combination of two vM distributions, as a concrete example of an angle distribution:
\begin{align} \label{DoublevM}
    g_{\kappa, \Delta\varphi} (\varphi_e) = q h_{\kappa, 0} (\varphi_e) + (1-q) h_{\kappa, \Delta\varphi} (\varphi_e),
\end{align}
where $q \in [0, 1]$ controls the mixedness between the two vM distributions, $\kappa$ controls the width of each peak, and $\Delta \varphi$ is the distance between the two peaks. For $q = 1/2$, we have $R = [I_2(\kappa) / I_0 (\kappa)]^2 \cos^2 (\Delta \varphi)$. Solving $R = R_c$, with $R_c$ being a critical point lying at the intersection of $\vec{n}$ and the phase boundaries in Fig.~\ref{fig:Setup_Summary}(b), one can obtain a phase diagram with respect to $\kappa$ and $\Delta \varphi$. Fig.~\ref{fig:Ex_DoublevMDist} shows the resulting phase diagram for the distribution Eq.~\eqref{DoublevM} with $q = 1/2$. Fig.~\ref{fig:Ex_DoublevMDist} illustrates that clear distinguishability between the two peaks (maximal at $\Delta \varphi =\pm \pi/2$) tends to degrade the quantum memory. It also breaks down for flat peaks (more precisely, $\kappa < 5.015$). The nontrivial point is that when $\vec{n}$ has sufficiently large $n_y$ so that it points within the green phase boundary, the quantum memory is robust regardless of the distinguishability between peaks and their width.

\subsection{Error Thresholds and Phase Boundary in the Replica Limit} \label{Sec:ETPBRL}

In the following, we discuss the practical error thresholds and mixed-state phase boundary in the replica limit under random rotation noise. 

In Appendix~\ref{App:Monotonicity}, we argue that the R\'enyi-2 mixed-state phase boundary $R_c^{(2)}$ gives an upper bound on the actual phase boundary $R_c^{(1)}$ assuming a physically reasonable flattening of the density matrix spectrum under decoherence. To assess how informative $R_c^{(2)}$, derived from the doubled Hilbert space formalism, is for practical QEC, we need to compare it with the actual phase boundary $R_c^{(1)}$ in the replica limit. For analytically tractable rotation axes, this comparison can be made directly: for $\vec{n} = \vec{x}, \vec{z}$, we have $R_c^{(2)} \simeq 0.586$ and $R_c^{(1)} \simeq 0.388$~\cite{dennis2002topological}, which are quite different. In contrast, for $\vec{n} = \hat{y}$, the two phase boundaries perfectly match at $R_c^{(2)} = R_c^{(1)} = 1$~\cite{tuckett2019tailoring}. Beyond these special cases, however, evaluating their closeness solely from the study based on the doubled Hilbert space formalism is challenging. 

To bridge this gap, we numerically estimate the error threshold $R_{\mathrm{th}}$ of the rotated surface code under stochastic $\vec{n}\cdot \vec{\sigma}$ noise ($\mathcal{E}_{\vec{n},p} [\rho] = (1-p) \rho + p (\vec{n} \cdot \vec{\sigma}) \rho (\vec{n} \cdot \vec{\sigma})$ with $p = (1 - \sqrt{1 - R})/2$). This is done using tensor network simulations~\cite{darmawan2017tensor}, which implement the optimal decoder for standard QEC scheme based on syndrome measurements. Note that the decohered surface code exhibits the same mixed-state transition point as the decohered toric code, as their corresponding stat-mech models differ only at the boundary, which does not affect bulk critical physics in the thermodynamic limit~\cite{niwa2025coherent}. We also chose the rotated surface code, whose layout achieves the optimal encoding ratio~\cite{bombin2007optimal}.

\begin{table*}[t]
    \centering
    \begin{tabular}{!{\vrule width 1.1pt}c|c!{\vrule width 1.1pt}c|c!{\vrule width 1.1pt}c|c!{\vrule width 1.1pt}}
        \noalign{\hrule height 1.1pt}   
        $\vec{n} (\phi, \theta)$ & $R_{\mathrm{th}}$ & $\vec{n} (\phi, \theta)$ & $R_{\mathrm{th}}$ & $\vec{n} (\phi, \theta)$ & $R_{\mathrm{th}}$ \\ 
        \noalign{\hrule height 1.1pt}   
        $\phi = \pi/2,\, \theta = \pi/2$ & $\simeq 0.38$ & $\phi = 3\pi/8,\, \theta = \pi/2$ & $\simeq 0.43$ & $\phi = \pi/4,\, \theta = \pi/2$ & $\simeq 0.68$ \\
        \hline
        $\phi = \pi/2,\, \theta = 3\pi/8$ & $\simeq 0.43$ & $\phi = 3\pi/8,\, \theta = 3\pi/8$ & $\simeq 0.47$ & $\phi = \pi/4,\, \theta = 3\pi/8$ & $\simeq 0.61$ \\
        \hline
        $\phi = \pi/2,\, \theta = \pi/4$ & $\simeq 0.67$ & $\phi = 3\pi/8,\, \theta = \pi/4$ & $\simeq 0.63$ & $\phi = \pi/4,\, \theta = \pi/4$ & $\simeq 0.60$ \\
        \hline
        $\phi = \pi/2,\, \theta = \pi/8$ & $\simeq 0.78$ & $\phi = 3\pi/8,\, \theta = \pi/8$ & $\simeq 0.74$ & $\phi = \pi/4,\, \theta = \pi/8$ & $\simeq 0.74$ \\
        \hline
        $\theta = 0$ & $1$ & & & & \\
        \noalign{\hrule height 1.1pt}   
    \end{tabular}
    \caption{Numerically estimated error thresholds $R_{\mathrm{th}}$ of the rotated surface code under random rotation noise for various rotation axes $\vec{n}$. The conventional QEC scheme based on syndrome measurement is simulated to compute $R_{\mathrm{th}}$.}
    \label{tab:threshold}
\end{table*}

\begin{figure*}[t]
    \centering
    \includegraphics[width=2\columnwidth]{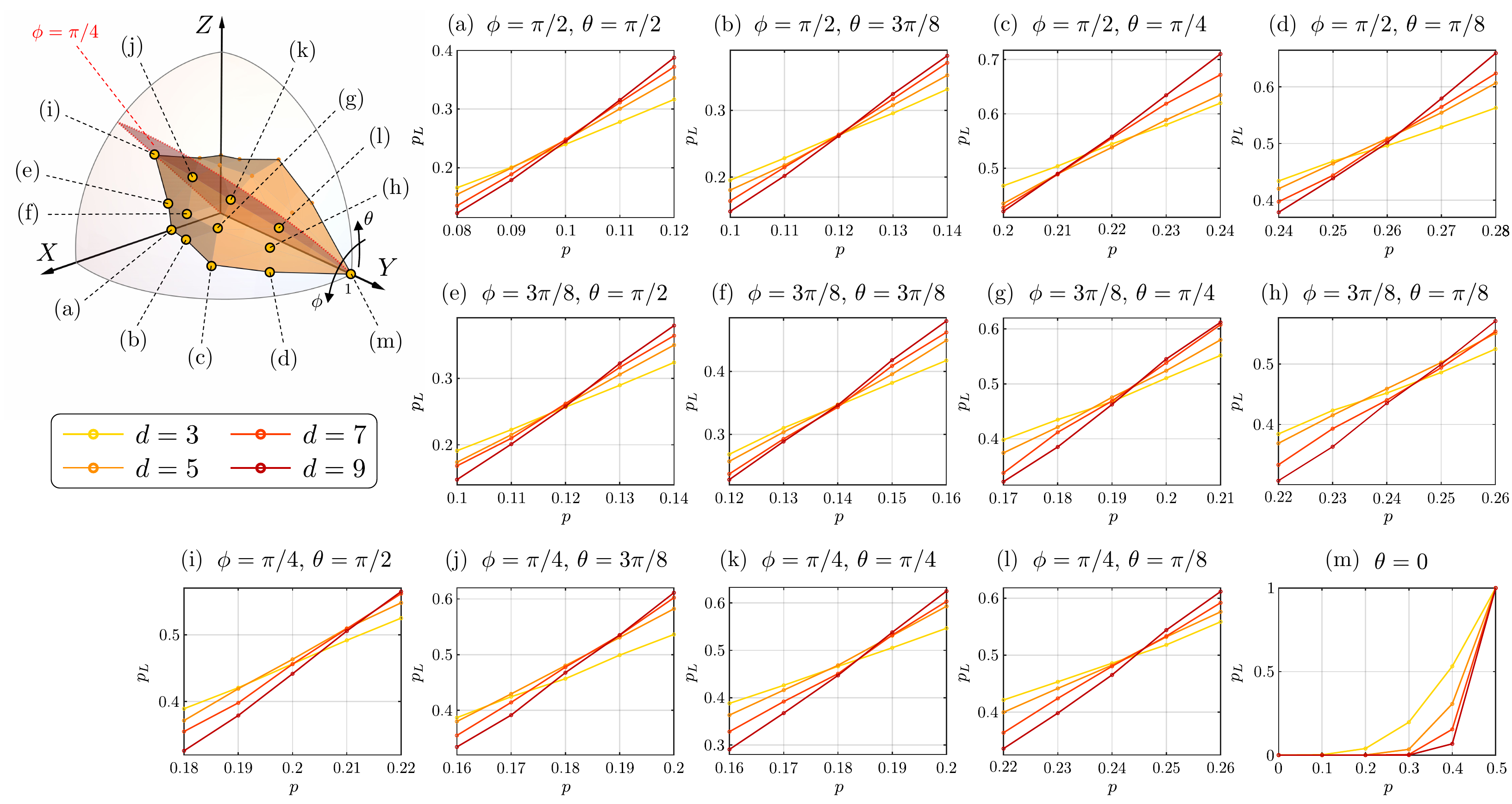}
    \caption{A three-dimensional illustration of the numerically estimated error thresholds $R_{\mathrm{th}}$ for the rotated surface code under random rotation noise. The thresholds above the $\phi = \pi/4$ plane (red) are obtained by reflecting the lower part. The threshold behavior used to extract $R_{\mathrm{th}}$ for each $\vec{n}$ is shown in (a)-(m) for code distances $d = 3, 5, 7, 9$. Each data point is obtained from $10^4$ to $4\times 10^4$ samples.}
    \label{fig:ErrThres}
\end{figure*}

For each rotation axis $\vec{n} = (\sin\theta \sin\phi, \cos\theta, \sin\theta \cos\phi)$ within an octant of the unit sphere, the estimated error thresholds $R_{\mathrm{th}}$ are listed in Table~\ref{tab:threshold}. (Thresholds for rotation axes in the other octants can be deduced by symmetry.) Figure~\ref{fig:ErrThres} provides a 3D illustration of $R_{\mathrm{th}}$ alongside the threshold behavior of the logical error rates $p_L$ as a function of the physical error rate $p$. (The logical error rate $p_L$ is defined as the diamond distance between the logical channel induced by QEC and the identity channel.) Though our numerical estimates are subject to finite-size effects, we expect these results to provide reasonably quantitative error thresholds under random rotation noise in the standard QEC scheme. 

The estimated error thresholds for the $\hat{x}$- and $\hat{z}$-axes are reasonably close to the known value $R_{\mathrm{th}} \simeq 0.388$ ~\cite{dennis2002topological} [see Fig.~\ref{fig:ErrThres}(a)]. Furthermore, we observe a trend where $R_{\mathrm{th}}$ increases as the rotation axis $\vec{n}$ approaches the $\hat{y}$-axis, reaching the maximal value $R_{\mathrm{th}} = 1$ for $\vec{n} = \hat{y}$, consistent with Ref.~\cite{tuckett2019tailoring} [see Fig.~\ref{fig:ErrThres}(m)]. Notably, these error thresholds can be achieved for arbitrary angle distributions by applying adequate single-qubit unitary rotations with a rotation angle given in Eq.~\eqref{varphi0} before syndrome measurements, which effectively reduces the noise to stochastic $\vec{n} \cdot \vec{\sigma}$ noise.

These error thresholds not only serve as practical benchmarks for the standard QEC scheme (with unitary pre-rotations before syndrome measurements allowed) but also provide a lower bound for the mixed-state phase boundary in the replica limit. This is because we cannot rule out the possibility that QEC protocols beyond the standard one might achieve a threshold higher than the obtained $R_{\mathrm{th}}$. Therefore, the actual mixed-state phase boundary in the replica limit lies somewhere between the estimated thresholds and the phase boundary obtained from the doubled Hilbert space formalism.

An intriguing point is that our phase diagram from the doubled Hilbert space formalism (and that in Ref.~\cite{chen2024unconventional}) suggests that quantum memory persists under random rotation noise up to maximal decoherence $R_c = 1$ for rotation axes $\vec{n}$ with $\phi = \pi/4$, whereas our numerical simulations of the standard QEC protocol do not seem to achieve such a high threshold [see Fig.~\ref{fig:ErrThres}(i)-(l)]. This discrepancy could be due to finite-size effects in our numerics or may indicate the existence of a non-standard QEC scheme capable of curing quantum memory up to $R_c = 1$. A precise determination of the mixed-state phase boundary in the replica limit and the construction of an explicit finite-depth local channel that attains it remain important open questions.
    
\section{Amplitude Damping Noise} \label{Sec:AmpDampNoise}

In this section, we study decohered TC under amplitude damping noise. The amplitude damping channel models the decay process of an excited state of a qubit via the spontaneous emission of a photon. Specifically, the channel is defined as $\EE_{\mathrm{amp}} \equiv \prod_e \EE_e$, where $\EE_e [\rho] \equiv K_{0,e} \rho K_{0,e}^\dagger + K_{1,e} \rho K_{1,e}^\dagger$ and the Kraus operators are given by
\begin{align}
    K_{0,e} = \begin{pmatrix}
        1 & 0 \\
        0 & \sqrt{1 - \gamma}
    \end{pmatrix}, \qquad K_{1,e} = \begin{pmatrix}
        0 & \sqrt{\gamma} \\
        0 & 0
    \end{pmatrix}.
\end{align}
The damping parameter $\gamma \in [0, 1]$ depends on the rate of spontaneous emission [see Fig.~\ref{fig:Setup_Summary}(a)]. The amplitude damping channel is a typical coherent noise in NISQ devices~\cite{nielsen2001quantum,chirolli2008decoherence,eczoo_ampdamp} and cannot be expressed as a stochastic Pauli channel. This channel is particularly relevant for quantum platforms consisting of lossy bosonic modes when the first two Fock states of a bosonic mode are used as a qubit, and it has been a noise model of great interest in the QEC literature~\cite{leung1997approximate,chuang1997bosonic,cochrane1999macroscopically,ouyang2014permutation,albert2018performance,jayashankar2022achieving}. In what follows, we are interested in the decohered state $\rho_D = \EE_{\mathrm{amp}} [\rho_0]$.

As in Sec.~\ref{Sec:MapSMModel}, we can compute the purity $\mathrm{Tr}[\rho_D^2] = \llangle \rho_0 | \bbE_{\mathrm{amp}}^\dagger \bbE_{\mathrm{amp}} | \rho_0 \rrangle$, where $\bbE_{\mathrm{amp}} = \prod_e \bbE_e$ is given by $\bbE_e = K_{0,e} \otimes \bar{K}_{0,e} + K_{1,e} \otimes \bar{K}_{1,e}$. In Appendix~\ref{App:AmpDampNoise}, we show that the purity can be written as $\Tr [\rho_D] \propto \sum_{\bfs_v, \bftau_v} \prod_e \omega_e$, where $s_v$ and $\tau_v$ are Ising spins on the vertices and the local weight is given by $\omega_e \propto 1 + J_+ s_v s_{v'} - J_- \tau_v \tau_{v'} + K s_v s_{v'} \tau_v \tau_{v'}$ with
\begin{align} \label{JK_AmpDamp}
    J_\pm = \frac{\gamma (1 \pm \gamma)}{\gamma^2 - \gamma + 2}, \qquad 
    K = \frac{\gamma^2 - 3\gamma + 2}{\gamma^2 - \gamma + 2}.
\end{align}
For $\gamma \in [0,1]$, $\omega_e$ with Eq.~\eqref{JK_AmpDamp} corresponds to the local Boltzmann weight of the anisotropic AT model (see Appendix~\ref{App:AmpDampNoise} for details). The phase diagram must be symmetric about $\gamma = 1/2$ since the stat-mech model is symmetric with respect to $\gamma \leftrightarrow 1 - \gamma$. (This transformation exchanges $J_+$ and $K$, which does not affect the partition function $\sum_{\bfs_v, \bftau_v} \prod_e \omega_e$ due to its invariance under the permutation of $s$ and $s\tau$ spins.) 

\begin{figure}[t]
    \includegraphics[width=\columnwidth]{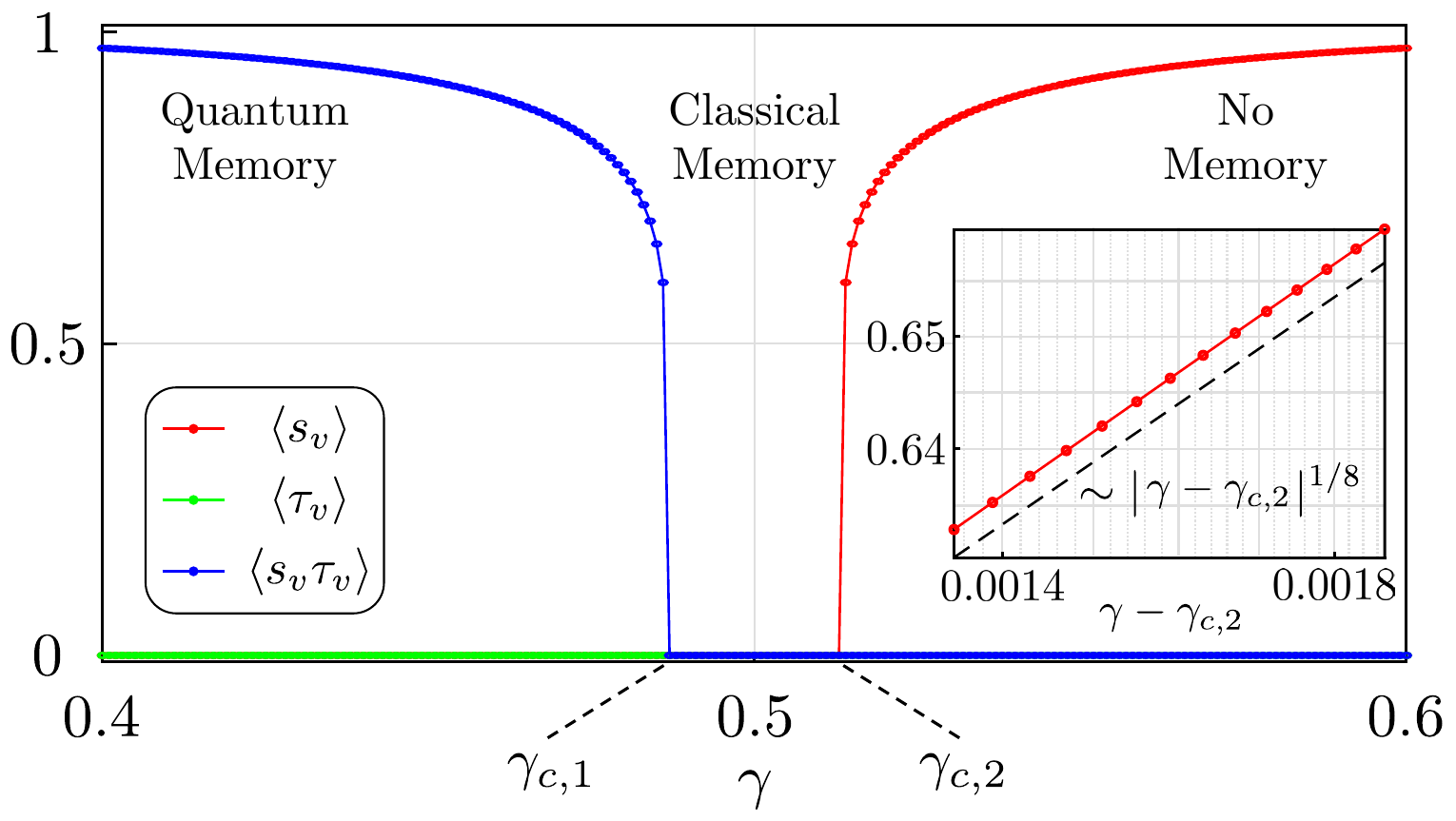}
    \caption{Order parameters $\langle s_v \rangle$, $\langle \tau_v \rangle$, and $\langle s_v \tau_v \rangle$ of the statistical mechanics model Eq.~\eqref{JK_AmpDamp} associated with the decohered toric code under amplitude damping noise. The model undergoes two phase transitions at $\gamma_{c,1} \simeq 0.487$ and $\gamma_{c,2} \simeq 0.513$, beyond each of which the quantum memory degrades to classical memory and no memory properties remain. The inset (log-log scale) shows the scaling of $\langle s_v \rangle$ above $\gamma_{c,2}$ with the Ising critical exponent $\beta = 1/8$.}
    \label{fig:CTMRG_AmpDamp}
\end{figure}

In the absence of decoherence ($\gamma = 0$), the weight reduces to $\omega_e \propto 1 + s_v s_{v'} \tau_v \tau_{v'}$, indicating that the model is in the PO phase with $\langle s_v \tau_v \rangle = 1$ and $\langle s_v \rangle = \langle \tau_v \rangle = 0$. On the other hand, when all spins are polarized to the ground state ($\gamma = 1$), the weight reduces to $\omega_e \propto 1 + s_v s_{v'}$ and hence the model has $\langle s_v \rangle = 1$ and $\langle \tau_v \rangle = \langle s_v \tau_v \rangle = 0$. Thus, one might expect only two phases: the PO phase at small $\gamma$, and the phase with ordered $s$ spins at large $\gamma$. In addition to these two phases, we find from the CTMRG simulation an intermediate PM phase with $\langle \tau_v \rangle = \langle s_v \rangle = \langle s_v \tau_v \rangle = 0$ in a narrow parameter regime $0.487 < \gamma < 0.513$ (see Fig.~\ref{fig:CTMRG_AmpDamp}). Therefore, the amplitude damping channel makes the decohered TC undergo two successive topological phase transitions, even though it has only one tuning parameter $\gamma$. This phenomenon has also been observed numerically in Ref.~\cite{li2025replica} using different schemes, and our work provides a solid stat-mech model underlying it.

The anyon parameters are identical to Eqs.~\eqref{AP_RandRot}, from which we can learn which anyons condense: The first phase transition at $\gamma_{c,1} \simeq 0.487$ condenses $m\bar{m}$ anyons and degrades the quantum memory to classical memory. Subsequently, the second phase transition at $\gamma_{c,2} \simeq 0.513$ condenses $e\bar{e}$ anyons, trivializing the classical memory. This coincides with almost all spins polarizing to the ground state for large $\gamma$. The overall phase diagram is shown in Fig.~\ref{fig:Setup_Summary}(c). The numerics show that the two transitions belong to the 2D Ising universality with critical exponent $\beta = 1/8$ (see inset of Fig.~\ref{fig:CTMRG_AmpDamp}). The R\'enyi-2 coherent information is also given by Eq.~\eqref{RandRotIc}, and an argument similar to the one in Sec.~\ref{Sec:GeneralPD} can be used to show that $I_c^{(2)}$ in the thermodynamic limit is equal to $2\log 2$, $0$, and $-2\log 2$ for the PO, PM, and the $s$-ordered phases, respectively. This is consistent with the analysis from the anyon parameters.

It is informative to compare our mixed-state phase boundary for amplitude damping noise with the error thresholds from Ref.~\cite{darmawan2017tensor}. There, the error threshold for amplitude damping noise in the standard QEC scheme is numerically estimated as $\gamma_c \simeq 0.39$, which is lower than $\gamma_{c,1} \simeq 0.487$ obtained from the doubled Hilbert space formalism. This highlights the formalism's limitation in pinpointing precise error thresholds. However, the doubled Hilbert space formalism provides additional physical insights into the mixed state that cannot be deduced from the numerical threshold estimate alone; it reveals a two-stage degradation of quantum memory---first to classical memory, then to a no-memory state---as $\gamma$ increases, while also identifying the physical mechanism behind each transition. Thus, complementing error threshold estimation with the R\'enyi-2 phase study offers a more comprehensive understanding of quantum memory under decoherence.

\section{Conclusions and Discussion} \label{Sec:Discussion}

In this work, we explore the mixed-state phase of the decohered toric code under two representative coherent noise using the doubled Hilbert space formalism. For random rotation noise, we determine the phase diagram for any rotation axis and arbitrary angle distribution, significantly generalizing the previously studied incoherent noise models. In addition to the robust quantum memory observed under random rotations about axes near the $Y$-axis, we identify phase boundaries featuring several interesting properties, such as Ising critical surfaces and unconventional mixed-state phase transitions, which merit further investigation. We argue that, under certain physical assumptions, the phase boundaries obtained from the doubled Hilbert space formalism provide an upper bound of the phase boundary in the replica limit. Additionally, we narrow down the location of the phase boundary in the replica limit by providing a numerical lower bound, which may be of practical interest. For the amplitude damping noise, we uncover two mixed-state topological phase transitions governed by a single parameter $\gamma$. 

We remark that the phase boundaries for the random rotation error identified here differ from the error threshold under global unitary rotation obtained in Ref.~\cite{venn2023coherent}, where the decoding problem maps to a Majorana network. This difference stems from practical constraints in standard QEC schemes, where syndromes are obtained by measuring check operators (e.g., $A_v$ and $B_p$) and Pauli recovery operations are applied. In contrast, our mixed-state setup addresses the intrinsic error threshold of the decohered toric code, within which QEC is possible \emph{in principle}, though the corresponding QEC protocol may require experimentally challenging operations. It would be interesting to analytically understand the ``practical'' error threshold for coherent noise, such as amplitude damping noise, as done for global rotation error~\cite{venn2023coherent}. Our finite-size numerics on random rotation noise hints at the approximate locations of such practical thresholds. The stat-mech mapping technique and tensor network method employed in this work would be straightforwardly applicable to yield more precise results in the thermodynamic limit.

As mentioned in Sec.~\ref{Sec:CriticalGreen}, it will be worthwhile to understand the precise conditions for an unconventional mixed-state phase transition. Moreover, an important open question is whether the structure of our mixed-state phase diagrams under coherent noise persists in the replica limit $n\rightarrow 1$. For coherent noise, exact diagonalization of the decohered density matrix, as done for incoherent noise~\cite{lee2025exact,niwa2025coherent}, is challenging. Developing new analytical and numerical frameworks to access the replica limit directly would be valuable for further study. 

Lastly, extending the mixed-state framework to other realistic QEC scenarios would be valuable. For example, realistic noise may introduce fluctuations in the rotation axis $\vec{n}$. Such fluctuations can be addressed similarly to our treatment of random rotation noise by incorporating a distribution over $\vec{n}$. It would be interesting to explore how these axis fluctuations influence the mixed-state phase diagram of our work. Furthermore, accounting for readout errors is crucial in practical QEC. In multi-cycle error detection setups, readout errors introduce an additional temporal error pattern~\cite{dennis2002topological}. Recently, Ref.~\cite{negari2024spacetime} has examined this scenario using a mixed-state perspective. Extending this approach to coherent noise would be an intriguing direction for future research.

\emph{Note added}: While completing this work, we became aware of related independent work that studies the error threshold of the surface code under generic unitary rotation errors using the mapping to a disordered non-Hermitian AT model~\cite{behrends2024statistical}.

\begin{acknowledgments}
    We thank Jong Yeon Lee and Tarun Grover for their helpful discussions. This work was supported by 2021R1A2C4001847, 2022M3H4A1A04074153, National Measurement Standard Services and Technical Services for SME, funded by the Korea Research Institute of Standards and Science (Grant No. KRISS–2024–GP2024-0015), Nano \& Material Technology Development Program through the National Research Foundation of Korea (NRF), funded by the Ministry of Science and ICT (Grant No. RS-2023-00281839), and Creation of the quantum information science R\&D ecosystem (based on human resources) through the NRF, funded by the Korean government (Ministry of Science and ICT (MSIT)) (Grant No. RS-2023-00256050).
\end{acknowledgments}

\section*{Data Availability}

The data supporting the findings of this article are available from the authors upon reasonable request.

\begin{appendix}
    \section{General Mapping for Anyon Parameters and Coherent Information}

    In this appendix, we review the mapping of anyon condensation/confinement parameters to an observable in the stat-mech model [Eqs.~\eqref{AnyonParamSM}], which was introduced in Ref.~\cite{chen2024unconventional}, for the sake of self-containedness. We then derive the stat-mech expression for the R\'enyi-2 coherent information [Eqs.~\eqref{MapIc} and \eqref{MapIcAT}]. 

    \subsection{Anyon Condensation/Confinement Parameters} \label{App:AnyonParam}
    
    Consider a state $\rho_0^{e\bar{e}} = w_e (l) \rho_0 w_e^\dagger (l)$, where $w_e (l) = \prod_{e\in l} Z_e$ is a string operator creating a pair of $e$ anyons at the endpoints $i$ and $j$ of an open string $l$ on the original lattice. From Eq.~\eqref{TCGS_Choi}, the associated Choi state $| \rho_0^{e\bar{e}} \rrangle = w_e (l) \bar{w}_e (l) | \rho_0 \rrangle$ is given by
    \begin{align} 
        &| \rho_0^{e\bar{e}} \rrangle \\
        &\propto \sum_{\substack{\bfz_e, \bar{\bfz}_e, \\ \bfx_v, \bar{\bfx}_v}} \prod_{e\in l} z_e \bar{z}_e \prod_e (1 + z_e x_v x_{v'}) (1 + \bar{z}_e \bar{x}_v \bar{x}_{v'}) | \bfz_e, \bar{\bfz}_e \rrangle \nonumber \\ 
        &= \sum_{\substack{\bfz_e, \bar{\bfz}_e, \\ \bfx_v, \bar{\bfx}_v}} x_i \bar{x}_i x_j \bar{x}_j \prod_e (1 + z_e x_v x_{v'}) (1 + \bar{z}_e \bar{x}_v \bar{x}_{v'}) | \bfz_e, \bar{\bfz}_e \rrangle, \nonumber
    \end{align}
    where we have used $z_e \bar{z}_e (1 + z_e x_v x_{v'}) (1 + \bar{z}_e \bar{x}_v \bar{x}_{v'}) = x_v x_{v'} \bar{x}_v \bar{x}_{v'} (1 + z_e x_v x_{v'}) (1 + \bar{z}_e \bar{x}_v \bar{x}_{v'})$ in the second line. Since the action of $\bbE$ does not affect the additional $x_i \bar{x}_i x_j \bar{x}_j$ factor, we obtain
    \begin{equation}
        \begin{aligned}
            \llangle I \bar{I} | e\bar{e} \rrangle &= \lim_{|i - j| \rightarrow \infty} \frac{\sum_{\bfx_v, \bar{\bfx}_v, \bft_v, \bar{\bft}_v} x_i \bar{x}_i x_j \bar{x}_j \prod_e \omega_e}{\sum_{\bfx_v, \bar{\bfx}_v, \bft_v, \bar{\bft}_v} \prod_e \omega_e} \\
            &= \lim_{|i-j| \rightarrow \infty} \langle x_i \bar{x}_i x_j \bar{x}_j \rangle,
        \end{aligned}
    \end{equation}
    where $\omega_e$ is given by Eq.~\eqref{weight}. Notice that the effect of applying $w_e (l)$ ($\bar{w}_e (l)$) on $|\rho_0 \rrangle$ is to add a factor $x_i x_j$ ($\bar{x}_i \bar{x}_j$). Using this, one can similarly show $\llangle e \bar{I} | e\bar{I} \rrangle = \lim_{|i-j| \rightarrow \infty} \langle x_i t_i x_j t_j \rangle$.

    Next, consider $\rho_0^{m\bar{m}} = w_m (\tilde{l}) \rho_0 w_m^\dagger (\tilde{l})$, where $w_m (\tilde{l}) = \prod_{m\in \tilde{l}} X_e$ is a string operator creating a pair of $m$ anyons at the endpoints $\tilde{i}$ and $\tilde{j}$ of an open string $\tilde{l}$ on the dual lattice. Since the $X_e$ operator flips $z_e$ to $-z_e$ in Eq.~\eqref{TCGS_Choi} (and similarly for $\bar{X}_e$), the Choi state $| \rho_0^{m\bar{m}} \rrangle = w_m (l) \bar{w}_m (l) | \rho_0 \rrangle$ is given by
    \begin{align} \label{rho0mm}
        | \rho_0^{m\bar{m}} \rrangle &\propto \sum_{\substack{\bfz_e, \bar{\bfz}_e, \\ \bfx_v, \bar{\bfx}_v}} \prod_e (1 + \eta_e z_e x_v x_{v'}) (1 + \eta_e \bar{z}_e \bar{x}_v \bar{x}_{v'}) | \bfz_e, \bar{\bfz}_e \rrangle,
    \end{align}
    where $\eta_e = -1$ if $e\in \tilde{l}$, and $1$ otherwise. It follows that
    \begin{equation}
        \begin{aligned}
            \llangle I \bar{I} | m\bar{m} \rrangle = \lim_{|\tilde{i} - \tilde{j}| \rightarrow \infty} \frac{\sum_{\bfx_v, \bar{\bfx}_v, \bft_v, \bar{\bft}_v} \prod_e \omega_e^\eta}{\sum_{\bfx_v, \bar{\bfx}_v, \bft_v, \bar{\bft}_v} \prod_e \omega_e},
        \end{aligned}
    \end{equation}
    where
    \begin{align} \label{weight_eta}
        &\omega_e^\eta \equiv \sum_{z_e, \bar{z}_e, z'_e, \bar{z}'_e = \pm 1} \llangle z'_e, \bar{z}'_e | \bbE_e^\dagger \bbE_e | \eta_e z_e, \eta_e \bar{z}_e \rrangle \\
        &\hspace{5pt} \times (1 + z_e x_v x_{v'}) (1 + \bar{z}_e \bar{x}_v \bar{x}_{v'}) (1 + z'_e t_v t_{v'}) (1 + \bar{z}'_e \bar{t}_v \bar{t}_{v'}). \nonumber
    \end{align}
    Note that the sign of interactions $x_v x_{v'}$ and $\bar{x}_v \bar{x}_{v'}$ are flipped along $\tilde{l}$ in Eq.~\eqref{weight_eta} compared to Eq.~\eqref{weight}. Therefore, we obtain $\llangle I \bar{I} | m\bar{m} \rrangle = \lim_{|\tilde{i} - \tilde{j}| \rightarrow \infty} \langle \mu_{\tilde{i}}^x \mu_{\tilde{i}}^{\bar{x}} \mu_{\tilde{j}}^x \mu_{\tilde{j}}^{\bar{x}} \rangle$, where $\mu^x$ and $\mu^{\bar{x}}$ are the disorder parameters of $x$ and $\bar{x}$ spins, respectively~\cite{kadanoff1971determination}. Observing that the effect of $w_m(\tilde{l})$ ($\bar{w}_m(\tilde{l})$) on $|\rho_0 \rrangle$ is to flip the sign of interactions $x_v x_{v'}$ ($\bar{x}_v \bar{x}_{v'}$) along $\tilde{l}$, one can similarly derive $\llangle m\bar{I} | m\bar{I} \rrangle = \lim_{|\tilde{i}-\tilde{j}| \rightarrow \infty} \langle \mu_{\tilde{i}}^x \mu_{\tilde{i}}^t \mu_{\tilde{j}}^x \mu_{\tilde{j}}^t \rangle$. This completes the proof of the correspondence Eq.~\eqref{AnyonParamSM}.

    \subsection{Coherent Information} \label{App:CI}

    The logical operators of the TC are given by $\bfZ_{1,2} \equiv \prod_{e\in l_{1,2}} Z_e$ and $\bfX_{1,2} \equiv \prod_{e\in \tilde{l}_{1,2}} X_e$, where $l_{1,2}$ ($\tilde{l}_{1,2}$) are two non-contractible loops on the original (dual) lattice. Let $| \Psi_0 \rangle$ be the ground state satisfying $\bfZ_{1,2} |\Psi_0 \rangle = | \Psi_0 \rangle$. Then, the four ground states can be written as $|\Psi_0^{a,b} \rangle \equiv \bfX_1^a \bfX_2^b | \Psi_0 \rangle$ with $a,b \in \{0, 1\}$. These ground states can be obtained from Eq.~\eqref{TCGS} by applying $\bfX_{1,2}$ properly, which flips $z_e$ to $-z_e$ for $e\in \tilde{l}_{1,2}$ in Eq.~\eqref{TCGS}.

    Now, we express the R\'enyi-2 coherent information $I_c^{(2)}$ in terms of the stat-mech model Eq.~\eqref{purity}. Let $\QQ$ be the TC system and $\RR$ be the two-qubit reference system, forming two Bell pairs with two logical qubits of $\QQ$. The density matrices for $\RR\QQ$ and $\QQ$ before channel $\EE$ are given by
    \begin{equation}
        \begin{aligned}
            \rho_{0, \RR\QQ} &= \frac 14 \sum_{\substack{a,b = 0,1 \\ a',b' = 0,1}} |a, b \rangle \langle a',b' |_\RR \otimes | \Psi_0^{a,b} \rangle \langle \Psi_0^{a',b'} |_\QQ, \\
            \rho_{0, \QQ} &= \frac 14 \sum_{a,b = 0,1} | \Psi_0^{a,b} \rangle \langle \Psi_0^{a,b} |_\QQ,
        \end{aligned}
    \end{equation}
    where $|a, b\rangle_R$ are orthonormal states in $\RR$. Let $\eta_e = -1$ ($\zeta_e = -1$) if $e \in l_1$ ($e \in l_2$) and $1$ otherwise. Using $\Tr[\rho^2] = \llangle \rho | \rho \rrangle$ and Eq.~\eqref{TCGS_Choi}, one can obtain
    \begin{align}
        \Tr [\EE [\rho_{0, \RR\QQ}]^2] &\propto \sum_{\substack{a,b = 0,1 \\ a',b' = 0,1}} \underbrace{\sum_{\bfx_v, \bar{\bfx}_v, \bft_v, \bar{\bft}_v} \prod_e \omega_e^{aba'b'}}_{\equiv\, \ZZ^{aba'b'}}, \\
        \Tr [\EE [\rho_{0, \QQ}]^2] &\propto \sum_{\substack{a,b = 0,1 \\ a',b' = 0,1}} \underbrace{\sum_{\bfx_v, \bar{\bfx}_v, \bft_v, \bar{\bft}_v} \prod_e \widetilde{\omega}_e^{aba'b'}}_{\equiv\, \widetilde{\ZZ}^{aba'b'}},
    \end{align}
    where 
    \begin{align} 
        &\omega_e^{aba'b'} \equiv \!\! \sum_{\substack{z_e, \bar{z}_e, \\ z'_e, \bar{z}'_e = \pm 1}} \!\! \llangle \eta_e^a \zeta_e^b z'_e, \eta_e^{a'} \zeta_e^{b'} \bar{z}'_e | \bbE_e^\dagger \bbE_e | \eta_e^a \zeta_e^b z_e, \eta_e^{a'} \zeta_e^{b'} \bar{z}_e \rrangle \nonumber \\
        &\hspace{5pt} \times (1 + z_e x_v x_{v'}) (1 + \bar{z}_e \bar{x}_v \bar{x}_{v'}) (1 + z'_e t_v t_{v'}) (1 + \bar{z}'_e \bar{t}_v \bar{t}_{v'}), \label{w1} \\
        &\widetilde{\omega}_e^{aba'b'} \equiv \!\! \sum_{\substack{z_e, \bar{z}_e, \\ z'_e, \bar{z}'_e = \pm 1}} \!\! \llangle \eta_e^a \zeta_e^b z'_e, \eta_e^a \zeta_e^b \bar{z}'_e | \bbE_e^\dagger \bbE_e | \eta_e^{a'} \zeta_e^{b'} z_e, \eta_e^{a'} \zeta_e^{b'} \bar{z}_e \rrangle \nonumber \\
        &\hspace{5pt} \times (1 + z_e x_v x_{v'}) (1 + \bar{z}_e \bar{x}_v \bar{x}_{v'}) (1 + z'_e t_v t_{v'}) (1 + \bar{z}'_e \bar{t}_v \bar{t}_{v'}), \label{w2}
    \end{align}
    The weights Eqs.~\eqref{w1} and \eqref{w2} are nothing but the original weight Eq.~\eqref{weight} with the sign of interactions flipped along non-contractible loops according to the labels $(a,b, a',b')$. Note that $\omega_e^{0000} = \widetilde{\omega}_e^{0000} = \omega_e$. Finally, the R\'enyi-2 coherent information becomes [see Eq.~\eqref{CI}]
    \begin{align} \label{MapIc}
        I_c^{(2)} (\RR\rangle \QQ) = \log \left( \frac{\sum_{a,b,a',b' = 0,1} e^{-\Delta F^{aba'b'}}}{\sum_{a,b,a',b' = 0,1} e^{-\Delta \widetilde{F}^{aba'b'}}} \right).
    \end{align}
    where $\Delta F^{aba'b'} \equiv - \log (\ZZ^{aba'b'} / \ZZ^{0000})$ is the free energy cost of forming defects along the non-contractible loops corresponding to $\omega_e^{aba'b'}$ (and similarly for $\Delta \widetilde{F}^{aba'b'}$).
    
    This expression for $I_c^{(2)}$ can be simplified further for the coherent noise considered in this paper, where the weight $\omega_e$ is of the Ashkin-Teller type: $\omega_e \propto 1 + J_1 s_v s_{v'} + J_2 \tau_v \tau_{v'} + K s_v s_{v'} \tau_v \tau_{v'}$, with $s_v = x_v \bar{x_v}$ and $\tau_v = \bar{x}_v t_v$ [see Eqs.~\eqref{SMModel}, \eqref{weightRandRot}, and \eqref{JK_AmpDamp}]. In such cases, Eqs.~\eqref{w1} and \eqref{w2} reduce to
    \begin{equation}
        \begin{aligned}
            \omega_e^{ab} &\propto 1 + \eta_e^a \zeta_e^b (J_1 s_v s_{v'} + J_2 \tau_v \tau_{v'}) + K s_v s_{v'} \tau_v \tau_{v'}, \\
            \widetilde{\omega}_e^{ab} &\propto 1 + J_1 s_v s_{v'} + \eta_e^a \zeta_e^b (J_2 \tau_v \tau_{v'} + K s_v s_{v'} \tau_v \tau_{v'}),
        \end{aligned}
    \end{equation}
    where we redefined labels as $a + a' \rightarrow a$ and $b + b' \rightarrow b$. Accordingly, Eq.~\eqref{MapIc} simplifies to Eq.~\eqref{RandRotIc}:
    \begin{align} \label{MapIcAT}
        I_c^{(2)} (\RR\rangle \QQ) = \log \left( \frac{\sum_{a,b = 0,1} e^{-\Delta F_{s,\tau}^{ab}}}{\sum_{a,b = 0,1} e^{-\Delta F_{\tau,s\tau}^{ab}}} \right),
    \end{align}
    where $\Delta F_{s,\tau}^{ab}$ is the free energy cost of forming defects in the interactions $s_v s_{v'}$ and $\tau_v \tau_{v'}$ along the non-contractible loop pattern $(a, b)$ (and similarly for $\Delta F_{s,s\tau}^{ab}$).
    
    \section{Details on Random Rotation Noise}

    In this appendix, we elaborate on the random rotation noise discussed in Sec.~\ref{Sec:RandRotNoise}. We prove the transformation of Eq.~\eqref{RandRotQC} into stochastic $\vec{n} \cdot \vec{\sigma}$ noise and the $R$-dependence of the effective stat-mech model. We also detail the mapping to the stat-mech model Eq.~\eqref{SMModel} and solve the staggered vertex model for the $Y$-rotation noise discussed in Sec.~\ref{Sec:RandYRot}.

    \subsection{Cases with Even Angle Distribution} \label{App:ReducStoNoise}

    Here, we show that random rotation channel Eq.~\eqref{RandRotQC} reduces to a stochastic $\vec{n} \cdot \vec{\sigma}$ noise when the distribution $g(\varphi_e)$ is even. We expand the global unitary rotation as 
    \begin{equation}
        \prod_e U_e(\varphi_e) = \sum_{\CC} A_-(\CC, \pmb{\varphi}) O_\CC^{(\vec{n})}
    \end{equation}
    where $\CC$ runs over all subsets, $O_\CC^{(\vec{n})} \equiv \prod_{e \in \CC} (\vec{n} \cdot \vec{\sigma})_e$, and
    \begin{equation}
        A_\pm (\CC, \pmb{\varphi}) \equiv \prod_{e\not\in \CC} \cos\varphi_e \prod_{e\in \CC} (\pm i \sin\varphi_e).
    \end{equation}
    The decohered state $\rho_D = \EE [\rho_0]$ can then be written as $\rho_D = \sum_{\CC, \bar{\CC}} \mathcal{A} (\CC, \bar{\CC}) O_\CC^{(\vec{n})} \rho_0 O_{\bar{\CC}}^{(\vec{n})}$ with $\mathcal{A} (\CC, \bar{\CC}) \equiv \int \prod_e d\varphi_e g(\varphi_e) A_- (\CC, \pmb{\varphi}) A_+ (\CC, \pmb{\varphi})$. When $g(\varphi_e)$ is even, the coefficient $\mathcal{A} (\CC, \bar{\CC})$ simplifies as follows:
    \begin{equation}
        \begin{aligned}
            \mathcal{A} (\CC, \bar{\CC}) &= \int \prod_e d\varphi_e g(\varphi_e) \prod_{e\not\in \CC, \bar{\CC}} \cos^2 \varphi_e \prod_{e\in \CC, \bar{\CC}} \sin^2 \varphi_e  \\
            &\hspace{7pt} \times \prod_{e\in \CC, e\not\in \bar{\CC}} \! (-i\cos\varphi_e \sin \varphi_e) \! \prod_{e\not\in \CC, e\in \bar{\CC}} \!  (i\cos\varphi_e \sin \varphi_e) \\
            &= \delta_{\CC, \bar{C}} \cdot (1-p)^{N - |\CC|} p^{|\CC|},
        \end{aligned}
    \end{equation}
    where we have defined $p \equiv \int_{\varphi_e} g(\varphi_e) \sin^2 \varphi_e$ and utilized $\int_{\varphi_e} g(\varphi_e) \sin \varphi_e \cos \varphi_e = 0$ in the second line. Thus, we have $\rho_D = \sum_\CC (1-p)^{N - |\CC|} p^{|\CC|} O_\CC^{(\vec{n})} \rho_0 O_\CC^{(\vec{n})}$, which is nothing but $\rho_0$ under the stochastic $\vec{n} \cdot \vec{\sigma}$ channel with probability $p$.

    \subsection{\texorpdfstring{$R$}{R}-Dependence of Mixed-State Phase} \label{App:RDependence}

    We first prove that general random rotation noise can always be transformed into stochastic $\vec{n}\cdot \vec{\sigma}$ noise and then discuss its consequence on the mixed-state phase under random rotation noise. Under the CJ isomorphism, the random rotation noise Eq.~\eqref{RandRotQC} with respect to the distribution $g (\varphi_e)$ is mapped to $\bbE_{\mathrm{rot}, g} = \prod_e \bbE_{g, e}$, with
    \begin{align} \label{bbEg}
        \bbE_{g, e} = \int _{\varphi_e} g(\varphi_e) e^{-i\varphi_e (\vec{n} \cdot \vec{\sigma})_e} \otimes \overline{e^{i\varphi_e (\vec{n}' \cdot \vec{\sigma})_e}}, 
    \end{align}
    where $\vec{n}' = (n_x, -n_y, n_z)$ and the subscript $g$ emphasizes that the noise is with respect to $g(\varphi_e)$. Using the identity $e^{\pm i\varphi_e (\vec{n} \cdot \vec{\sigma})_e} = \cos\varphi_e I_e \pm i \sin\varphi_e (\vec{n} \cdot \vec{\sigma})_e$, we obtain
    \begin{equation}
        \begin{aligned}
            \mathbb{E}_{g,e} &= \frac 12 \left[ I \otimes \bar{I} + (\vec{n} \cdot \vec{\sigma}) \otimes \overline{(\vec{n}' \cdot \vec{\sigma})} \right]_e \\
            &\hspace{11pt} + \frac{r_c(g)}{2} \left[ I \otimes \bar{I} - (\vec{n} \cdot \vec{\sigma}) \otimes \overline{(\vec{n}' \cdot \vec{\sigma})} \right]_e \\
            &\hspace{11pt} + \frac{i r_s(g)}{2} \left[ I \otimes \overline{(\vec{n}' \cdot \vec{\sigma})} - (\vec{n} \cdot \vec{\sigma}) \otimes \bar{I} \right]_e,
        \end{aligned}
    \end{equation}
    where the parameters
    \begin{align}
        r_c(g) &\equiv \int_{\varphi_e} g(\varphi_e) \cos 2\varphi_e, \\
        r_s(g) &\equiv \int_{\varphi_e} g(\varphi_e) \sin 2\varphi_e
    \end{align}
    satisfy the relation $r_c^2(g) + r_s^2(g) = 1 - R$, with
    \begin{align}
        R &\equiv 1 - \left| \int_{\varphi_e} g(\varphi_e) e^{2i\varphi_e} \right|^2.
    \end{align}
    This shows that the random rotation channel depends on the distribution $g(\varphi)$ only through $r_c(g)$ and $r_s(g)$, which lie on a circle of radius $\sqrt{1 - R}$. Now, note that applying a rotation $e^{-i \varphi_0 (\vec{n} \cdot \vec{\sigma})}$ rotates the point $(r_c(g), r_s(g))$ on this circle counterclockwise by an angle $2\varphi_0$, i.e., 
    \begin{align}
        \begin{pmatrix}
            r_c(\tilde{g}) \\ r_s(\tilde{g})
        \end{pmatrix} = \begin{pmatrix}
            \cos 2\varphi_0 & -\sin 2\varphi_0 \\
            \sin 2\varphi_0 & \cos 2\varphi_0
        \end{pmatrix} \begin{pmatrix}
            r_c(g) \\ r_s(g)
        \end{pmatrix},
    \end{align}
    where $\tilde{g} (\varphi) \equiv g(\varphi - \varphi_0)$ is the original distribution shifted by $\varphi_0$. From this observation, we can always set $r_s(\tilde{g}) = 0$ by choosing an angle $\varphi_0$ given by 
    \begin{align} \label{varphi0}
        \varphi_0 \equiv -\frac 12 \tan^{-1} \left( \frac{r_s(g)}{r_c(g)} \right),
    \end{align}
    in which case the remaining parameter $r_c(\tilde{g})$ becomes $\sqrt{1 - R}$ and the channel simplifies to the stochastic $\vec{n} \cdot \vec{\sigma}$ channel
    \begin{align}
        \mathbb{E}_{\tilde{g}, e} = (1-p) I_e \otimes \bar{I}_e + p (\vec{n} \cdot \vec{\sigma})_e \otimes \overline{(\vec{n}' \cdot \vec{\sigma})_e}
    \end{align}
    with an error rate $p \equiv (1 - r_c(\tilde{g})) / 2 = (1 - \sqrt{1 - R}) / 2$. In other words, regardless of the distribution $g(\varphi)$, we can always transform random rotation noise with rotation axis $\vec{n}$ into stochastic $\vec{n} \cdot \vec{\sigma}$ noise by applying a depth-1 unitary rotation deterministically. Extending to the tensor product of $\bbE_{g,e}$ is straightforward. The cases with an even distribution discussed in Appendix~\ref{App:ReducStoNoise} correspond to the special case where $r_s(g) = 0$ and thus $\varphi_0 = 0$, which does not require any unitary rotation.

    The above transformation implies that states decohered under random rotation noise with different distributions belong to the same mixed-state phase as long as they share the same rotation axis $\vec{n}$ and the same $R$ value. This follows from the fact that the mixed-state phase remains unchanged under a finite-depth local channel. Furthermore, it directly follows that all $n$th moments of the decohered density matrix, $\Tr[\rho_D^n]$, depend on the distribution $g(\varphi_e)$ solely through a single parameter $R$.
    
    \begin{widetext}
        \subsection{Details on Mapping to Statistical Mechanics Model} \label{App:StatMechMap}

        Here, we fill in the details of the mapping to the stat-mech model Eq.~\eqref{SMModel} discussed in Section.~\ref{Sec:MapSMModel} by assuming even distribution $g(\varphi_e)$. We start by nothing that $\bbE_{g_1, e} \bbE_{g_2, e} = \bbE_{g_1 * g_2, e}$, where $\mathbb{E}_{g_{1,2},e}$ is given in Eq.~\eqref{bbEg} and $(g_1 * g_2) (\varphi_e) \equiv \int_{\varphi'_e} g_1 (\varphi'_e) g_2 (\varphi_e - \varphi'_e)$ is a convolution between the two distributions $g_1$ and $g_2$. Since $\bbE_{g,e}^\dagger = \bbE_{g_-, e}$ with $g_- (\varphi_e) \equiv g(-\varphi_e)$, the purity of $\rho_D$ can be written as
        \begin{align} \label{RhoEfkgRho}
            \Tr [\rho_D^2] = \llangle \rho_0 | \bbE_{\mathrm{rot},g}^\dagger \bbE_{\mathrm{rot},g} | \rho_0 \rrangle = \llangle \rho_0 | \bbE_{\mathrm{rot}, \fkg} | \rho_0 \rrangle,
        \end{align}
        where $\fkg \equiv g_- * g$. Note that $\fkg = g_- * g$ is even when $g$ is even. Let $\langle \cdot \rangle_{\fkg}$ denote the average with respect to $\fkg$. Using $\langle \sin \varphi \cos\varphi \rangle_{\fkg} = 0$, we can expand $\bbE_{\mathrm{rot}, \fkg}$ as follows:
        \begin{equation} \label{Efkg}
            \begin{aligned} 
                \bbE_{\mathrm{rot}, \fkg} &\propto \prod_e \big[ 1 + \lambda (n_x^2 X_e \bar{X}_e - n_y^2 Y_e \bar{Y}_e + n_z^2 Z_e \bar{Z}_e) \\
                &\hspace{30pt} + \lambda n_x n_y (Y_e \bar{X}_e - X_e \bar{Y}_e) + \lambda n_y n_z (Y_e \bar{Z}_e - Z_e \bar{Y}_e) + \lambda n_z n_x (Z_e \bar{X}_e + X_e \bar{Z}_e) \big], 
            \end{aligned}
        \end{equation}
        where $\lambda \equiv \langle \sin^2 \varphi \rangle_{\fkg} / \langle \cos^2 \varphi \rangle_{\fkg}$. Let's first analyze the range of $\lambda$. We have $\lambda = (1 - 2\pi a_2) / (1 + 2\pi a_2)$ with $a_2 \equiv (2\pi)^{-1} \int_{\varphi_e} \fkg (\varphi_e) e^{2i\varphi_e}$ being the second Fourier coefficient of $\fkg$, which must be non-negative by the convolution theorem. Combining this with $|a_2| \leq (2\pi)^{-1}$, we obtain $\lambda \in [0, 1]$.
        
        Now, we argue that the six cross-terms in Eq.~\eqref{Efkg} (e.g., $Y_e \bar{X}_e, Z_e \bar{X}_e, \dots$) can be neglected in the thermodynamic limit. Consider sandwiching $\bbE_{\mathrm{rot}, \fkg}$ with $\llangle \rho_0 |$ and $| \rho_0 \rrangle$. Note that $\llangle \rho_0 | \prod_{e \in l} Z_e \bar{Z}_e | \rho_0 \rrangle = 1$ ($\llangle \rho_0 | \prod_{e \in \tilde{l}} X_e \bar{X}_e | \rho_0 \rrangle = 1$) only when $l$ ($\tilde{l}$) forms closed loops in the original lattice $\LL$ (dual lattice $\DD$). Meanwhile, for $Y_e \bar{Y}_e$ and the six cross-terms in Eq.~\eqref{Efkg} (denoting these operators by $A_e \bar{B}_e$), we have $\llangle \rho_0 | \prod_{e \in l} A_e \bar{B}_e | \rho_0 \rrangle = 1$ only if $l$ forms closed loops along diagonals of the lattice. All the other terms appearing in the expansion of the product Eq.~\eqref{Efkg} vanish. From these properties of the TC, the purity $\Tr [\rho_D^2] = \llangle \rho_0 | \bbE_{\mathrm{rot}, \fkg} | \rho_0 \rrangle$ can be written as
        \begin{align}
            \Tr [\rho_D^2] = \langle \cos^2 \varphi \rangle_{\fkg}^N \cdot \left( \ZZ_1^{(\LL)} + \ZZ_2^{(\LL)} \right),
        \end{align}
        where the $\ZZ_1^{(\LL)}$ is the sum of all non-vanishing terms consisting of non-cross terms ($1$, $X_e \bar{X}_e$, $Y_e \bar{Y}_e$, $Z_e \bar{Z}_e$), and $\ZZ_2^{(\LL)}$ is the sum of all other non-vanishing terms containing at least one cross-term. The superscript $(\LL)$ indicates that the terms are with respect to the lattice $\LL$. Now, observe that
        \begin{align}
            \left| \frac{\ZZ_2^{(\LL)}}{\ZZ_1^{(\LL)}} \right| &\leq \sum_{l = 1}^{N_d} 6^l \sum_{S \in \SSS_l} \Lambda^{|S|} \cdot \left| \frac{\ZZ_1^{(\LL \setminus S)}}{\ZZ_1^{(\LL)}} \right|,
        \end{align}
        where $\Lambda \equiv \lambda \cdot \max\{ n_x n_y, n_y n_z, n_z n_x \}$. Here, $N_d$ is the number of diagonal loops on $\LL$, $\SSS_l$ is the set of subsets of qubits on $\LL$ that are unions of $l$ diagonal loops (with qubits appearing even times excluded), and $\LL \setminus S$ is the lattice obtained from $\LL$ by removing qubits in $S \in \SSS_l$. Note that $|S| \sim L_d l$ for $S \in \SSS_l$, where $L_d$ is the length of the shortest diagonal loop. Since all terms in $\ZZ_1^{(\LL \setminus S)}$ are included in $\ZZ_1^{(\LL)}$, we can write $\ZZ_1^{(\LL)} = \ZZ_1^{(\LL \setminus S)} + \ZZ_{\text{rem}}$ for some remainder $\ZZ_{\text{rem}}$. Note that all non-vanishing terms in $\ZZ_1^{(\LL)}$ are non-negative since the lengths of the diagonal loops are even. Then, we have
        \begin{align} \label{ZZRatio}
            \left| \frac{\ZZ_1^{(\LL \setminus S)}}{\ZZ_1^{(\LL)}} \right| = \left| \frac{1}{1 + \ZZ_{\text{rem}} / \ZZ_1^{(\LL \setminus S)} } \right| \leq 1,
        \end{align}
        and consequently
        \begin{align}
            \left| \frac{\ZZ_2^{(\LL)}}{\ZZ_1^{(\LL)}} \right| &\lesssim \sum_{l = 1}^{N_d} \binom{N_d}{l} (6\Lambda^{-L_d})^l = (1 + 6\Lambda^{-L_d})^{N_d} - 1 = \OO (\text{poly}(L) e^{-L}),
        \end{align}
        since $L_d = \OO(L)$ in general and $\max\{ n_x n_y, n_y n_z, n_z n_x \} \leq 1/2$ under the constraint $|\vec{n}| = 1$ so that $\Lambda \leq 1/2$. [When the size of $\LL$ is $L_x \times L_y$ with $\gcd(L_x, L_y) = 1$, one gets $L_d = L^2$, which greatly enhances the suppression.] Therefore, $\Tr [\rho_D^2] \propto \ZZ_1^{(\LL)}$ as $L\rightarrow \infty$, justifying that the six cross-terms in Eq.~\eqref{Efkg} can be safely neglected when discussing the phase of the model. 
        
        Now that $\Tr [\rho_D^2] \propto \llangle \rho_0 | \! \prod_e \! \left[ 1 + \lambda (n_x^2 X_e \bar{X}_e - n_y^2 Y_e \bar{Y}_e + n_z^2 Z_e \bar{Z}_e) \right] \!\! | \rho_0 \rrangle$, we can use the identities $X_e |z_e \rangle = |{-}z_e\rangle$, $Y_e |z_e \rangle = i^{z_e} |{-}z_e\rangle$ and $Z_e |z_e \rangle = z_e | z_e\rangle$ to yield
        \begin{equation}
            \begin{aligned}
                \Tr [\rho_D^2] &\propto \sum_{\bfz_e, \bar{\bfz}_e} \sum_{\bfx_v, \bar{\bfx}_v} \sum_{\bft_v, \bar{\bft}_v} \prod_e (1 + z_e t_v t_{v'}) (1 + \bar{z}_e \bar{t}_v \bar{t}_{v'}) \\
                &\hspace{15pt} \times \left[ (1 + \lambda n_z^2 z_e \bar{z}_e) (1 + z_e x_v x_{v'}) (1 + \bar{z}_e \bar{x}_v \bar{x}_{v'}) + \lambda (n_x^2 - i^{-z_e - \bar{z}_e} n_y^2) (1 - z_e x_v x_{v'}) (1 - \bar{z}_e \bar{x}_v \bar{x}_{v'}) \right] \\
                &\propto \sum_{\bfx_v, \bar{\bfx}_v, \bft_v, \bar{\bft}_v} \prod_e \omega_e,
            \end{aligned}
        \end{equation}
        with
        \begin{align} \label{weRandRot}
            \omega_e &\propto 1 + x_v x_{v'} \bar{x}_v \bar{x}_{v'} t_v t_{v'} \bar{t}_v \bar{t}_{v'}  + \frac{\lambda n_z^2}{1 + \lambda n_x^2} (x_v x_{v'} \bar{t}_v \bar{t}_{v'} + t_v t_{v'} \bar{t}_v \bar{t}_{v'} + \bar{x}_v \bar{x}_{v'} t_v t_{v'} + x_v x_{v'} \bar{x}_v \bar{x}_{v'}) \\
            &\hspace{10pt} + \frac{1 - \lambda n_x^2}{1 + \lambda n_x^2} (x_v x_{v'} t_v t_{v'} + \bar{x}_v \bar{x}_{v'} \bar{t}_v \bar{t}_{v'}) + \frac{\lambda n_y^2}{1 + \lambda n_x^2} (x_v x_{v'} \bar{x}_v \bar{x}_{v'} + t_v t_{v'} \bar{t}_v \bar{t}_{v'} - x_v x_{v'} \bar{t}_v \bar{t}_{v'} - \bar{x}_v \bar{x}_{v'} t_v t_{v'}). \nonumber
        \end{align}
        Since Eq.~\eqref{weRandRot} obeys $\omega_e (x_v x_{v'} \bar{x}_v \bar{x}_{v'} t_v t_{v'} \bar{t}_v \bar{t}_{v'}) = \omega_e$, we can let $\bar{t}_v \bar{t}_{v'} = x_v x_{v'} \bar{x}_v \bar{x}_{v'} t_v t_{v'}$. Introducing new Ising variables $s_v = x_v \bar{x}_v$ and $\tau_v = \bar{x}_v t_v$ and substituting $\lambda = R / (2-R)$ (which can be derived from $R = 2 \langle \sin^2 \varphi \rangle_{\fkg}$), we finally obtain Eqs.~\eqref{SMModel} and \eqref{weightRandRot}:
        \begin{align} \label{weRandRot2}
            \omega_e \propto 1 + \frac{R (n_z^2 + n_y^2)}{2 - R + R n_x^2} s_v s_{v'} + \frac{R (n_z^2 - n_y^2)}{2 - R + R n_x^2} \tau_v \tau_{v'} + \frac{2 - R - R n_x^2}{2 - R + R n_x^2} s_v s_{v'} \tau_v \tau_{v'}.
        \end{align}
        Writing Eq.~\eqref{weRandRot2} as a local Boltzmann weight $\omega_e \propto e^{\JJ_1  s_v s_{v'} + \JJ_2 \tau_v \tau_{v'} + \KK s_v s_{v'} \tau_v \tau_{v'}}$, it must be
        \begin{equation}
            \begin{aligned}
                \frac{\tanh(\JJ_1) + \tanh(\JJ_2) \tanh(\KK)}{1 + \tanh(\JJ_1) \tanh(\JJ_2) \tanh(\KK)} &= \frac{R (n_z^2 + n_y^2)}{2 - R + R n_x^2}, \\
                \frac{\tanh(\JJ_2) + \tanh(\JJ_1) \tanh(\KK)}{1 + \tanh(\JJ_1) \tanh(\JJ_2) \tanh(\KK)} &= \frac{R (n_z^2 - n_y^2)}{2 - R + R n_x^2}, \\
                \frac{\tanh(K) + \tanh(\JJ_1) \tanh(\JJ_2)}{1 + \tanh(\JJ_1) \tanh(\JJ_2) \tanh(\KK)} &= \frac{2 - R - R n_x^2}{2 - R + R n_x^2},
            \end{aligned}
        \end{equation}
        whose solution is imaginary in general. Therefore, the purity of the TC under random rotation noise is proportional to the partition function of the non-Hermitian anisotropic AT model.
    \end{widetext}
    
    \subsection{Staggered Vertex Model} \label{App:StagVertexModel}

\begin{figure}[b]
        \includegraphics[width=\columnwidth]{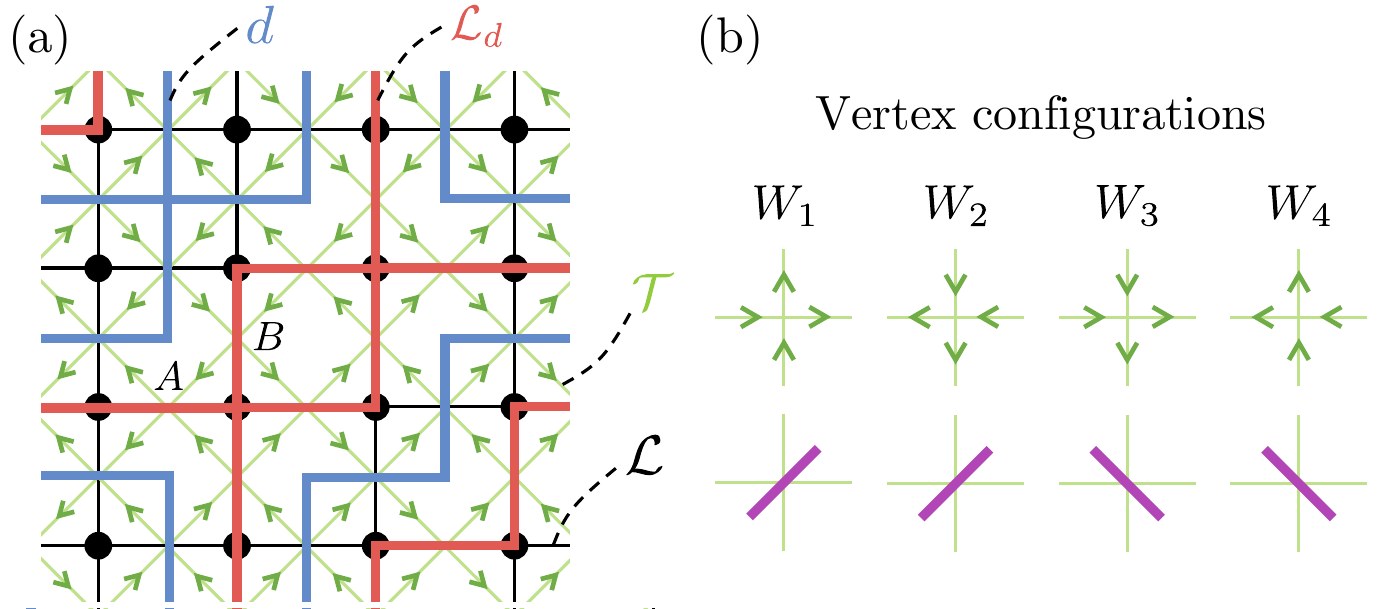}
        \caption{(a) Example of loop configuration $d$ (thick blue lines) on the dual lattice $\DD$ such that $\LL_d$ (thick red lines) is also a loop configuration. Black lines represent the original lattice $\LL$. Black dots represent the sites where $s$ and $\tau$ spins live. Green lines represent the tilted lattice $\TT$. Green arrows show one of the associated vertex configurations. (b) Each line segment diagonal in $\TT$ (purple segments in the bottom line, which can be segments of either $d$ or $\LL_d$) maps to the vertex configurations in the middle line with the corresponding weights $W_i$ ($1\leq i\leq 4$). Once a possible vertex is selected in one vertex, the other vertices are uniquely determined according to this assignment.}
        \label{fig:Mapping_StagVertexModel}
    \end{figure}

    Here, we derive and solve the staggered vertex model (defined below) for the random $Y$-rotation noise discussed in Sec.~\ref{Sec:RandYRot}. Let's start from Eq.~\eqref{Weight_Y}. We find it convenient to exchange $s$ spins and $s\tau$ spins and then flip the $\tau$ spins located at one sublattice (say, $\tau_{(i,j)}$ with $i+j$ odd), so that
    \begin{align} \label{YWeight}
        \omega_e \propto 1 + \frac{R}{2 - R} s_v s_{v'} (1 + \tau_v \tau_{v'}) - \tau_v \tau_{v'}.
    \end{align}
    (The partition function is invariant under such transformations.) Following an approach akin to Ref.~\cite{nienhuis1984critical, saleur1987partition}, we map the non-Hermitian AT model Eq.~\eqref{YWeight} to a certain vertex model as follows. Note that $\omega_e = [2R / (2 - R)] s_v s_{v'}$ if $\tau_v \tau_{v'} = 1$, and 2 otherwise. Let $\LL$ and $\DD$ denote the original and dual lattices, respectively. Then, the partition function $\ZZ = \sum_{\bfs_v, \bftau_v} \prod_e \omega_e$ can be expanded as
    \begin{equation}
        \begin{aligned}
            \ZZ &\propto \sum_{d \subset \DD} \left( \frac{R}{2 - R} \right)^{N - |d|} \sum_{\bfs_v} \prod_{e\in \LL_d} s_v s_{v'} \\
            &\propto \sum'_{d\subset \DD} \left( \frac{R}{2 - R} \right)^{N - |d|},
        \end{aligned}
    \end{equation}
    where $d$ runs over all closed loop configurations on the dual lattice $\DD$ (which represent edges with $\tau_v \tau_{v'} = -1$), $|d|$ is the total length of the loop configuration $d$, and $\LL_d$ is a lattice obtained from $\LL$ by removing edges intersecting with $d$. In the second line, we used the fact that $\sum_{\bfs_v} \prod_{e\in \LL_d} s_v s_{v'} = 2^{N/2}$ only when all vertices of $\LL_d$ have even number of edges (i.e., $\LL_d$ is itself a loop configuration) and vanishes otherwise. Here, $'$ indicates that the sum is over all loop configurations $d \subset \DD$ such that $\LL_d$ is also a loop configuration. An example of such configuration $d$ is shown in Fig.~\ref{fig:Mapping_StagVertexModel}(a). 
    
    Let $\mathcal{T}$ be a $45^\circ$-tilted square lattice whose vertices pass across all the edges of $\mathcal{L}$. Let $A$ ($B$) be the sublattice of $\mathcal{T}$ that horizontal edges of $\mathcal{L}$ ($\mathcal{D}$) pass through. Now, $\ZZ$ can be mapped to a certain vertex model via the following weight assignment:
    \begin{equation} \label{VertexWeight}
        \begin{aligned}
            W_1^A = W_2^A = W_3^B = W_4^B &= 1, \\
            W_3^A = W_4^A = W_1^B = W_2^B &= \frac{R}{2 - R},
        \end{aligned}
    \end{equation}
    where the superscripts denote the sublattices in $\mathcal{T}$, and the weights $W_i$ ($1\leq i\leq 4$) are assigned to the local vertices as shown in Fig.~\ref{fig:Mapping_StagVertexModel}(b). (This mapping is one-to-two: two vertex configurations with opposite arrows map to the same $d$.) The weights Eq.~\eqref{VertexWeight} vary depending on the sublattice and hence define a ``staggered vertex'' model. When $R = 1$, the staggeredness disappears, reducing the model to the ``right-angle water'' ice model~\cite{ziman1979models}. [For a finite torus geometry, some loop configuration can lead to antiperiodic boundary conditions along the non-contractible loops of the torus, i.e., the arrows change their directions across these loops~\cite{saleur1987partition}. Since we are interested in the thermodynamic limit, we consider an infinite plane geometry and neglect such an issue below.]

    \begin{figure}[b]
        \includegraphics[width=\columnwidth]{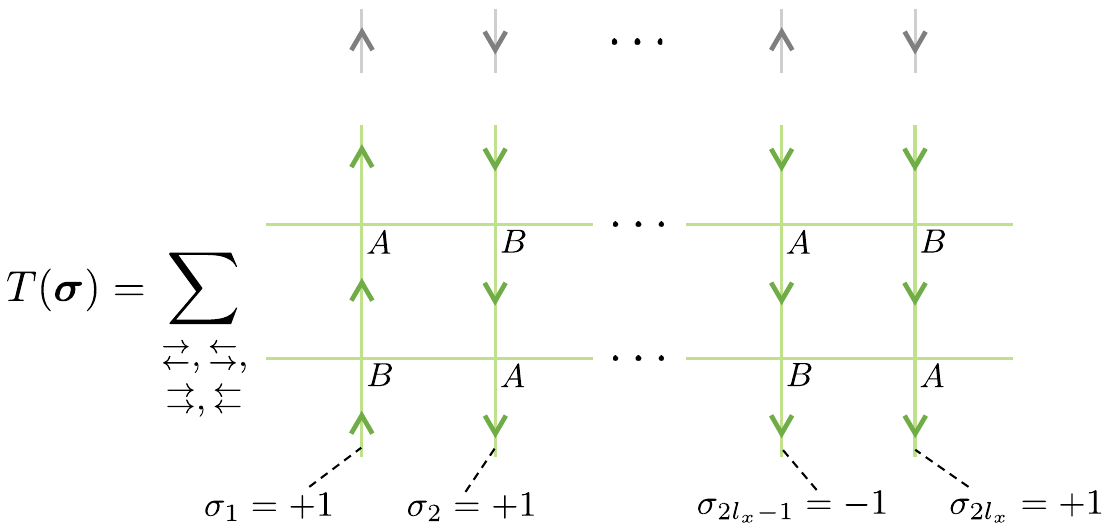}
        \caption{Transfer matrix element $T(\bfsig)$ of the staggered vertex model. Gray arrows at the top show the fiducial antiferromagnetic arrow pattern, based on which variables $\bfsig = (\sigma_i)_{i=1}^{2l_x}$ are defined. Each $T(\bfsig)$ is a sum of the four terms corresponding to the four-row patterns $\substack{\leftarrow\\[-1em] \rightarrow}$, $\substack{\rightarrow\\[-1em] \leftarrow}$, $\substack{\rightarrow\\[-1em] \rightarrow}$, and $\substack{\leftarrow\\[-1em] \leftarrow}$.}
        \label{fig:YRot_TransMat}
    \end{figure}

    The model Eq.~\eqref{VertexWeight} can be easily solved due to the non-reversing arrows along each line of $\TT$ [see Fig.~\ref{fig:Mapping_StagVertexModel}(a)]. Let's first compute the transfer matrix of the model Eq.~\eqref{VertexWeight} in the diagonal direction of the original lattice $\LL$, i.e., along the direction of the tilted lattice $\TT$. Let's suppose that $\TT$ has a size $2l_x \times 2l_y$ for a while, and then take a limit $l_{x,y} \rightarrow \infty$ to compute physical quantities in an infinite plane geometry. Define variables $\bfsig = (\sigma_i)_{i=1}^{2l_x}$ along the $l_x$-direction as follows (see Fig.~\ref{fig:YRot_TransMat}): Consider a fiducial antiferromagnetic arrow pattern $\uparrow \downarrow \uparrow \downarrow \cdots$, which starts from a site in the $A$ sublattice. Define $\sigma_i = 1$ ($-1$) if the vertical arrow of the $i$th site matches (mismatches) that of the fiducial pattern. Let's also consider a small alternating electric field along the fiducial pattern, i.e., upward for the odd lines and downward for the even lines, so that the vertices with vertical arrows aligning with (against) the electric field gain an additional $\epsilon$ ($\epsilon^{-1}$) factor for some $\epsilon > 1$. We consider this electric field to consider a spontaneous symmetry breaking, choosing one of the two vertical arrow patterns satisfying $O(\bfsig) \equiv \sum_{i=1}^{2l_x} \sigma_i = \pm 2l_x$.

    Let $r = R / (2 - R)$ and $T$ be a transfer matrix corresponding to two rows, which involve four cases for their arrow directions: $\substack{\leftarrow\\[-1em] \rightarrow}$, $\substack{\rightarrow\\[-1em] \leftarrow}$, $\substack{\rightarrow\\[-1em] \rightarrow}$, and $\substack{\leftarrow\\[-1em] \leftarrow}$ (see Fig.~\ref{fig:YRot_TransMat}). Due to the non-reversing arrows, we can write $T = \sum_{\bfsig} T(\bfsig) | \bfsig \rangle \langle \bfsig |$ with $T(\bfsig) = \prod_{i=1}^{2l_x} t(\sigma_i)$, where $t(\sigma_i)$ is a local weight for the $i$th column. One can compute $t(\sigma_i)$ for the four cases as follows:
    \begin{enumerate}[leftmargin=*]
        \item Case $\substack{\leftarrow\\[-1em] \rightarrow}$: \,$t(\sigma_i) = \epsilon^{2\sigma_i} r^{1 - \sigma_i}$
        \begin{itemize}[leftmargin=*]
            \item When $\sigma_i = 1$ for odd $i$, $t(\sigma_i) = W_1^A W_4^B = \epsilon^2$.

            \item When $\sigma_i = 1$ for even $i$, $t(\sigma_i) = W_3^B W_2^A = \epsilon^2$.

            \item When $\sigma_i = -1$ for odd $i$, $t(\sigma_i) = W_3^A W_2^B = \epsilon^{-2} r^2$.
            
            \item When $\sigma_i = -1$ for even $i$, $t(\sigma_i) = W_1^B W_4^A = \epsilon^{-2} r^2$. 
        \end{itemize}
        
        \item Case $\substack{\rightarrow\\[-1em] \leftarrow}$: \,$t(\sigma_i) = \epsilon^{2\sigma_i} r^{1 + \sigma_i}$
        \begin{itemize}[leftmargin=*]
            \item When $\sigma_i = 1$ for odd $i$, $t(\sigma_i) = W_4^A W_1^B = \epsilon^2 r^2$. 

            \item When $\sigma_i = 1$ for even $i$, $t(\sigma_i) = W_2^B W_3^A = \epsilon^2 r^2$. 

            \item When $\sigma_i = -1$ for odd $i$, $t(\sigma_i) = W_2^A W_3^B = \epsilon^{-2}$. 
            
            \item When $\sigma_i = -1$ for even $i$, $t(\sigma_i) = W_4^B W_1^A = \epsilon^{-2}$.
        \end{itemize}
        
        \item Case $\substack{\rightarrow\\[-1em] \rightarrow}$: \,$t(\sigma_i) = \epsilon^{2\sigma_i} r$
        \begin{itemize}[leftmargin=*]
            \item When $\sigma_i = 1$ for odd $i$, $t(\sigma_i) = W_1^A W_1^B = \epsilon^2 r$.

            \item When $\sigma_i = 1$ for even $i$, $t(\sigma_i) = W_3^B W_3^A = \epsilon^2 r$.

            \item When $\sigma_i = -1$ for odd $i$, $t(\sigma_i) = W_3^A W_3^B = \epsilon^{-2} r$.
            
            \item When $\sigma_i = -1$ for even $i$, $t(\sigma_i) = W_1^B W_1^A = \epsilon^{-2} r$.
        \end{itemize}
        
        \item Case $\substack{\leftarrow\\[-1em] \leftarrow}$: \,$t(\sigma_i) = \epsilon^{2\sigma_i} r$
        \begin{itemize}[leftmargin=*]
            \item When $\sigma_i = 1$ for odd $i$, $t(\sigma_i) = W_4^A W_4^B = \epsilon^2 r$.

            \item When $\sigma_i = 1$ for even $i$, $t(\sigma_i) = W_2^B W_2^A = \epsilon^2 r$.

            \item When $\sigma_i = -1$ for odd $i$, $t(\sigma_i) = W_2^A W_2^B = \epsilon^{-2} r$.
            
            \item When $\sigma_i = -1$ for even $i$, $t(\sigma_i) = W_4^B W_4^A = \epsilon^{-2} r$.
        \end{itemize}
    \end{enumerate}
    Combining all, we obtain
    \begin{align} \label{T}
        T(\bfsig) = 2 \epsilon^{O(\bfsig)} r^{2l_x} \left[ 1 + \cosh \left( O(\bfsig) \log r \right) \right].
    \end{align}
    Note that $T$ only depends on $O(\bfsig)$, whose value takes even integers from $-2l_x$ to $2l_x$. Since $T(\bfsig)$ is monotonic in $|O(\bfsig)|$ when $\epsilon = 1$, the correlation length is given by
    \begin{align} \label{xi}
        \xi = -\frac{1}{\log \left( \frac{1 + \cosh [2(l_x-1) \log r]}{1 + \cosh [2l_x \log r]} \right)} \underset{l_x \rightarrow \infty}{\longrightarrow} -\frac{1}{2 \log r},
    \end{align}
    which diverges as $[4(1-R)]^{-1}$ as $R \rightarrow 1$.
    
    The order parameter $O = \langle O(\bfsig) \rangle / (2l_x)$ can be directly computed from Eq.~\eqref{T}. Consider first a case with $0 \leq R < 1$. It is straightforward to show that
    \begin{align}
        O &= \frac{\sum_{\bfsig} \frac{1}{2l_x} O(\bfsig) T(\bfsig)^{l_y}}{\sum_{\bfsig} T(\bfsig)^{l_y}} \\
        &= \frac{\frac{1}{l_x} \sum_{p = -l_x}^{l_x} p \binom{2l_x}{l_x + p} \epsilon^{2 l_y p} \left[ 1 + \cosh \left( 2p \log r \right) \right]^{l_y}}{\sum_{p = -l_x}^{l_x} \binom{2l_x}{l_x + p} \epsilon^{2 l_y p} \left[ 1 + \cosh \left( 2p \log r \right) \right]^{l_y}}, \nonumber
    \end{align}
    where we have let $2p = O(\bfsig)$. Since $\epsilon > 1$, the expression $\epsilon^{2 l_y p} \left[ 1 + \cosh \left( 2p \log r \right) \right]^{l_y}$ dominates at $p = l_x$ as $l_y \rightarrow \infty$. Therefore, we have
    \begin{align}
        O &\underset{l_y \rightarrow \infty}{\longrightarrow} \frac{\frac{1}{l_x} \cdot l_x \cdot \epsilon^{2 l_x l_y} \left[ 1 + \cosh \left( 2l_x \log r \right) \right]^{l_y}}{\epsilon^{2 l_x l_y} \left[ 1 + \cosh \left( 2l_x \log r \right) \right]^{l_y}} = 1.
    \end{align}
    For the case of $R = 1$, $T(\bfsig) = 4\epsilon^{O(\bfsig)}$ and hence
    \begin{align}
        O &= \frac{\frac{1}{l_x} \sum_{p = -l_x}^{l_x} p \binom{2l_x}{l_1 + p} \epsilon^{2 l_y p}}{\sum_{p = -l_x}^{l_x} \binom{2l_x}{l_x + p} \epsilon^{2 l_y p}} = \frac{\epsilon^{2l_y} - 1}{\epsilon^{2l_y} + 1}.
    \end{align}
    One can observe that $\lim_{\epsilon \rightarrow 1^+} \lim_{l_{x,y} \rightarrow \infty} O = 1$, while $\lim_{l_{x,y} \rightarrow \infty} \lim_{\epsilon \rightarrow 1^+} O = 0$. This is the spontaneous breaking of the symmetry that reverses the arrow directions. This behavior of the order parameter $O$ and the diverging correlation length at $R = 1$ indicate that the model Eq.~\eqref{VertexWeight} does not undergo phase transition for $R < 1$.

    \begin{figure}[b]
        \includegraphics[width=0.96\columnwidth]{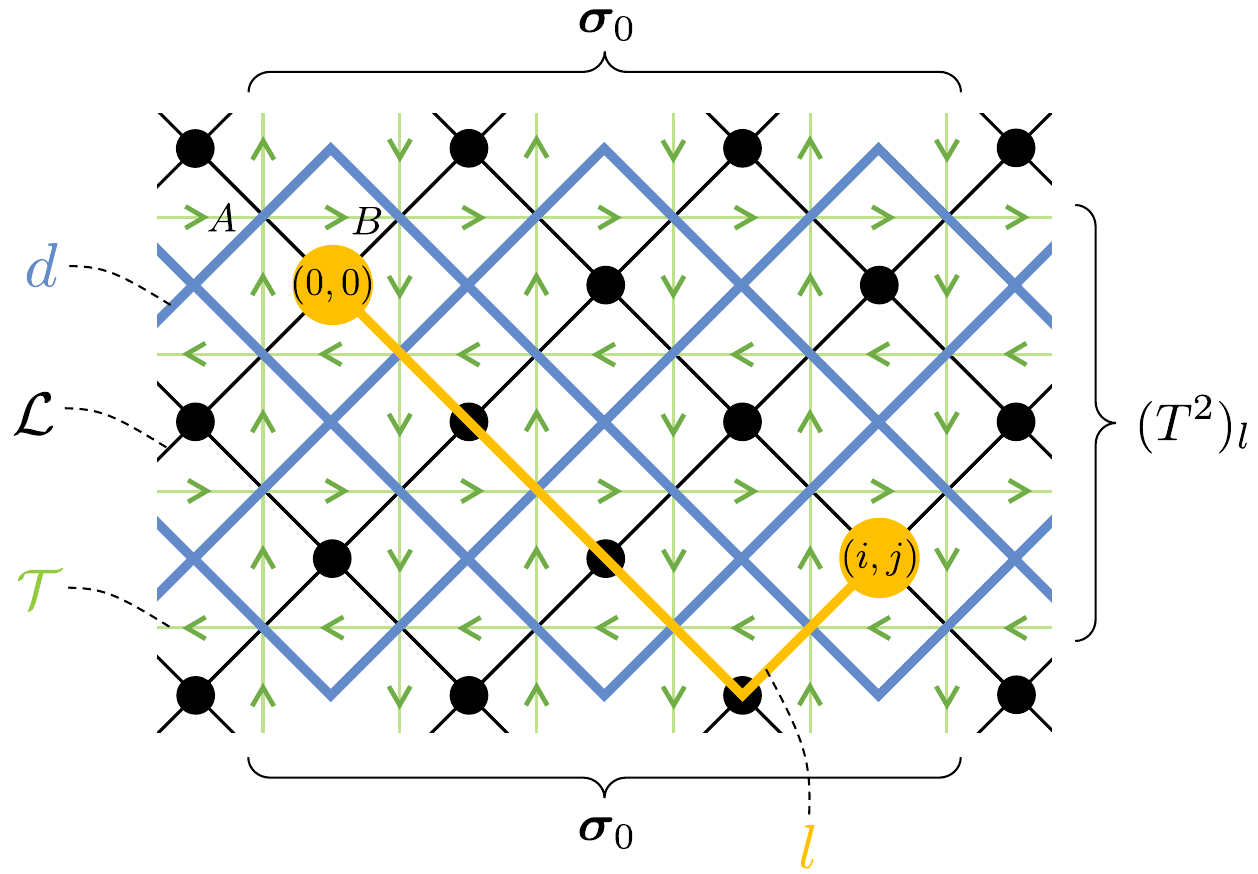}
        \caption{Modified transfer matrix element $\langle \bfsig_0 | (T^q)_l | \bfsig_0 \rangle$ with $q = 2$ appearing in the computation of $\langle \tau_{(0,0)} \tau_{(i,j)} \rangle$ in the staggered vertex model Eq.~\eqref{VertexWeight}. The dominant loop configuration that survives under the limit $l_x \rightarrow \infty$ (blue loops) and the corresponding arrow pattern are displayed. Vertices where the yellow string passes through have flipped vertex weight when it intersects with $d$, yielding a $(-1)^{i+j}$ factor.}
        \label{fig:YRot_AnyonParam}
    \end{figure}

    Now, we compute the anyon parameters for $R < 1$, whose stat-mech expressions follow from Eq.~\eqref{AnyonParamSM}:
    \begin{equation} 
        \begin{aligned}
            \llangle I\bar{I} | e\bar{e} \rrangle &= \lim_{|i-j| \rightarrow \infty} \langle s_i s_j \rangle, \\
            \llangle e\bar{I} | e\bar{I} \rrangle &= \lim_{i^2 + j^2 \rightarrow \infty} (-1)^{i+j} \langle \tau_{(0,0)} \tau_{(i,j)} \rangle, \\
            \llangle I\bar{I} | m\bar{m} \rrangle &= \lim_{\tilde{i}^2 + \tilde{j}^2 \rightarrow \infty} (-1)^{\tilde{i}+\tilde{j}} \langle \mu_{(\tilde{0}, \tilde{0})}^\tau \mu_{(\tilde{i}, \tilde{j})}^\tau \rangle, \\
            \llangle m\bar{I} | m\bar{I} \rrangle &= \lim_{|\tilde{i}-\tilde{j}| \rightarrow \infty} \langle \mu_{\tilde{i}}^s \mu_{\tilde{j}}^s \rangle,
        \end{aligned}
    \end{equation}
    [This differs from Eq.~\eqref{AP_RandRot} since we have exchanged $s$ and $s\tau$ spins and then flipped the $\tau$ spins placed at site $(i,j)$ with $i + j$ odd.] Let's first compute $\langle \tau_{(0,0)} \tau_{(i,j)} \rangle = \langle \prod_{e\in l} \tau_v \tau_{v'} \rangle$, where $l$ is any string on the original lattice $\LL$ connecting sites $(0,0)$ and $(i,j)$. From Eq.~\eqref{YWeight}, we have $w_e \tau_i \tau_j = [2R/(2-R)] s_v s_{v'}$ if $\tau_v \tau_{v'} = -1$, and $-2$ otherwise. This corresponds to flipping $1$ to $-1$ in Eq.~\eqref{VertexWeight} for the vertices passing through edges $e \in l$. Suppose that $l$ is contained in $2q$ rows of $\TT$ (see Fig.~\ref{fig:YRot_AnyonParam} for $q = 2$). Then, since $| \bfsig_0 \rangle$ with $\bfsig_0 = (\sigma_i = 1)_{i=1}^{2l_x}$ is a eigenvector of $T$ with a dominant eigenvalue, we can write $\langle \tau_{(0,0)} \tau_{(i,j)} \rangle = \langle \bfsig_0 | (T^q)_l | \bfsig_0 \rangle / \langle \bfsig_0 | T^q | \bfsig_0 \rangle$ as $l_y \rightarrow \infty$, where $(T^q)_l$ is the same as $T^q$ except for flipping $1$ to $-1$ in Eq.~\eqref{VertexWeight} for the vertices passing through the intersections between $l$ and loops $d$ on dual lattice $\DD$. Now, as $l_x \rightarrow \infty$, one can easily see that only the loop pattern full of $d$ (see Fig.~\ref{fig:YRot_AnyonParam}) survives (other patterns give powers of $x^{2l_x}$, which vanish as $l_x \rightarrow \infty$). Since $l$ intersects with this loop pattern $i+j$ times, one get $\langle \bfsig_0 | (T^q)_l | \bfsig_0 \rangle = (-1)^{i+j} \langle \bfsig_0 | T^q | \bfsig_0 \rangle$ and therefore
    \begin{align} \label{Y_eIeI}
        \llangle e\bar{I} | e\bar{I} \rrangle = \lim_{i^2 + j^2 \rightarrow \infty} (-1)^{i+j} \langle \tau_{(0,0)} \tau_{(i,j)} \rangle = 1.
    \end{align}
    Following a similar argument, one can see that $\langle \mu_{\tilde{i}}^s \mu_{\tilde{j}}^s \rangle = \langle \bfsig_0 | (T^q)_{\tilde{l}} | \bfsig_0 \rangle / \langle \bfsig_0 | T^q | \bfsig_0 \rangle$ as $l_y \rightarrow \infty$, where $\tilde{l}$ is a string on the dual lattice $\DD$ connecting sites $\tilde{i}$ and $\tilde{j}$, and $(T^q)_{\tilde{l}}$ is the same as $T^q$ except for flipping $r$ to $-r$ in Eq.~\eqref{VertexWeight} for the vertices passing through edges $e \in \tilde{l}$. Again, the loop pattern shown in Fig.~\ref{fig:YRot_AnyonParam} survives under the limit $l_x \rightarrow \infty$, yielding $\langle \bfsig_0 | (T^q)_{\tilde{l}} | \bfsig_0 \rangle = \langle \bfsig_0 | T^q | \bfsig_0 \rangle$ and hence
    \begin{align} \label{Y_mImI}
        \llangle m\bar{I} | m\bar{I} \rrangle = \lim_{|i - j| \rightarrow \infty} \langle \mu_{\tilde{i}}^s \mu_{\tilde{j}}^s \rangle = 1.
    \end{align}
    One can also show that the fermion condensation parameter $\llangle f\bar{I} | f\bar{I} \rrangle = (-1)^{i+j} \langle \tau_{(0,0)} \mu_{(\tilde{0}, \tilde{0})}^s \tau_{(i,j)} \mu_{(\tilde{i}, \tilde{j})}^s \rangle$ converges to one in the same manner.
    
    For the anyon condensation parameters, one can borrow the theorem in Ref.~\cite{levin2020constraints} stating that order and disorder parameters of Ising symmetric system with finite correlation length cannot be nonzero simultaneously. Since the correlation length Eq.~\eqref{xi} is finite for $R < 1$ and the disorder parameters Eqs.~\eqref{Y_eIeI} and \eqref{Y_mImI} are finite, we can conclude that the corresponding order parameters must vanish, i.e., $\llangle I\bar{I} | e\bar{e} \rrangle = \llangle m\bar{I} | m\bar{I} \rrangle = 0$. 

    To sum up, we have shown that $e$ and $m$ anyons are deconfined and $e\bar{e}$ and $m\bar{m}$ anyons are not condensed for all $R < 1$. Along with the analysis of correlation length and order parameter $O$, it follows that the staggered vertex model does not undergo phase transition for $R < 1$. 

    \begin{widetext}
        \section{Details on Amplitude Damping Noise} \label{App:AmpDampNoise}
    
        In this appendix, we elaborate on the details of the mapping to the stat-mech model discussed in Section.~\ref{Sec:AmpDampNoise}. We start by computing $\bbE_{\mathrm{amp}}^\dagger \bbE_{\mathrm{amp}}$ for the amplitude damping channel. From $\bbE_e = K_{0,e} \otimes \bar{K}_{0,e} + K_{1,e} \otimes \bar{K}_{1,e}$, one can easily show that
        \begin{align} \label{EE_AmpDamp}
            \bbE_{\mathrm{amp}}^\dagger \bbE_{\mathrm{amp}} \propto \prod_e \Bigg[ 1 + \frac{\gamma (1 - \gamma)}{\gamma^2 - 2\gamma + 2} (Z_e + \bar{Z}_e) + \frac{1}{\gamma^2 - 2\gamma + 2} \left( \gamma^2 Z_e \bar{Z}_e + \gamma X_e \bar{X}_e - \gamma Y_e \bar{Y}_e \right) \Bigg].
        \end{align}
        From Eq.~\eqref{TCGS_Choi} and the identities $X_e |z_e \rangle = |{-}z_e\rangle$, $Y_e |z_e \rangle = i^{z_e} |{-}z_e\rangle$, and $Z_e |z_e \rangle = z_e | z_e\rangle$, it follows that $\Tr [\rho_D^2] = \llangle \rho_0 | \bbE_{\mathrm{amp}}^\dagger \bbE_{\mathrm{amp}} | \rho_0 \rrangle \propto \sum_{\bfx_v, \bar{\bfx}_v, \bft_v, \bar{\bft}_v} \prod_e \omega_e$, where
        \begin{align} \label{weAmpDamp}
            \omega_e &= \sum_{z_e, \bar{z}_e = \pm 1} (1 + z_e t_v t_{v'}) (1 + \bar{z}_e \bar{t}_v \bar{t}_{v'}) \nonumber  \\
            &\hspace{15pt} \times \left[ \left(1 + \frac{\gamma^2 z_e \bar{z}_e + \gamma ( 1-\gamma) (z_e + \bar{z}_e)}{\gamma^2 - 2\gamma + 2} \right) (1 + z_e x_v x_{v'}) (1 + \bar{z}_e \bar{x}_v \bar{x}_{v'}) + \frac{\gamma (1 - i^{-z_e - \bar{z}_e})}{\gamma^2 - 2\gamma + 2} (1 - z_e x_v x_{v'}) (1 - \bar{z}_e \bar{x}_v \bar{x}_{v'}) \right] \nonumber \\
            &\propto 1 + x_v x_{v'} \bar{x}_v \bar{x}_{v'} t_v t_{v'} \bar{t}_v \bar{t}_{v'} + \frac{\gamma^2 - 3\gamma + 2}{\gamma^2 - \gamma + 2} (x_v x_{v'} t_v t_{v'} + \bar{x}_v \bar{x}_{v'} \bar{t}_v \bar{t}_{v'}) + \frac{\gamma (1 + \gamma)}{\gamma^2 - \gamma + 2} (x_v x_{v'} \bar{x}_v \bar{x}_{v'} + t_v t_{v'} \bar{t}_v \bar{t}_{v'}) \\
            &\hspace{12pt} + \frac{\gamma (1 - \gamma)}{\gamma^2 - \gamma + 2} (x_v x_{v'} + \bar{x}_v \bar{x}_{v'} + t_v t_{v'} + \bar{t}_v \bar{t}_{v'} - x_v x_{v'} \bar{t}_v \bar{t}_{v'} - \bar{x}_v \bar{x}_{v'} t_v t_{v'} \nonumber \\
            &\hspace{135pt} + x_v x_{v'} \bar{x}_v \bar{x}_{v'} t_v t_{v'} + x_v x_{v'} t_v t_{v'} \bar{t}_v \bar{t}_{v'} + x_v x_{v'} \bar{x}_v \bar{x}_{v'} \bar{t}_v \bar{t}_{v'} + \bar{x}_v \bar{x}_{v'} t_v t_{v'} \bar{t}_v \bar{t}_{v'}). \nonumber
        \end{align}
        Since Eq.~\eqref{weAmpDamp} obeys $\omega_e (x_v x_{v'} \bar{x}_v \bar{x}_{v'} t_v t_{v'} \bar{t}_v \bar{t}_{v'}) = \omega_e$, we can let $\bar{t}_v \bar{t}_{v'} = x_v x_{v'} \bar{x}_v \bar{x}_{v'} t_v t_{v'}$ and obtain 
        \begin{align} \label{weAmpDamp2}
            \omega_e &\propto 1 + \frac{\gamma (1 + \gamma)}{\gamma^2 - \gamma + 2} CA - \frac{\gamma (1 - \gamma)}{\gamma^2 - \gamma + 2} BC + \frac{\gamma^2 - 3\gamma + 2}{\gamma^2 - \gamma + 2} AB + \frac{\gamma (1 - \gamma)}{\gamma^2 - \gamma + 2} (A + B + C + ABC),
        \end{align}
        where we simplified the notations as $A \equiv x_v x_{v'}$, $B = t_v t_{v'}$, and $C \equiv \bar{x}_v \bar{x}_{v'}$. 
        
        We argue that we can neglect the last term in Eq.~\eqref{weAmpDamp2} proportional to $A + B + C + ABC$. When $\gamma = 0$ or $1$, the last term of Eq.~\eqref{weAmpDamp2} vanishes trivially. Now, consider the case $0 < \gamma < 1$. Let's first represent Eq.~\eqref{weAmpDamp2} as a local Boltzmann weight. To this end, it is convenient to consider the cases with $B = \pm C$ separately.
        \begin{enumerate}[leftmargin=*]
            \item When $B = C$, Eq.~\eqref{weAmpDamp2} reduces to
            \begin{align} \label{BCp1}
                \omega_e \propto \frac{2(\gamma^2 - \gamma + 1)}{\gamma^2 - \gamma + 2} \left[ 1 + AB + \frac{\gamma (1 - \gamma)}{\gamma^2 - \gamma + 1} (A + B) \right] = \frac{2(\gamma^2 - \gamma + 1)}{\gamma^2 - \gamma + 2} \frac{\sech^2 (c_1) \sech(c_2)}{1 + \tanh^2 (c_1) \tanh(c_2)} e^{c_1 (A+B) + c_2 AB},
            \end{align}
            where $c_1 = \tanh^{-1} \left[ (\gamma^2 - \gamma + 1 - \sqrt{2\gamma^2 - 2\gamma + 1}) / \gamma (1 - \gamma) \right] < \infty$ and $c_2 = \infty$. For a while, we formally regard $c_2$ as a large number. 
            
            \item When $B = -C$, Eq.~\eqref{weAmpDamp2} reduces to
            \begin{align} \label{BCm1}
                \omega_e \propto \frac{2}{\gamma^2 - \gamma + 2} \left[ 1 - (1 - 2\gamma) AB \right] = \frac{4 \sqrt{\gamma (1 - \gamma)}}{\gamma^2 - \gamma + 2} e^{-c_3 AB},
            \end{align}
            where $c_3 = \tanh^{-1} (1 - 2\gamma) < \infty$.
        \end{enumerate}
        Define
        \begin{align}
            c_4 = \log \left( \frac{\gamma^2 - \gamma + 1}{2 \sqrt{\gamma (1 - \gamma)}} \frac{\sech^2 c_1 \, \sech c_2}{1 + \tanh^2 c_1 \tanh c_2} \right).
        \end{align}
        Since $e^{\frac{c_4}{2} BC}$ becomes the prefactors of Eqs.~\eqref{BCp1} and \eqref{BCm1} (up to some common prefactor) for $B = \pm C$, respectively, we can combine these two cases using projectors $(1 \pm BC) / 2$ as 
        \begin{align} \label{weAmpDamp3}
            \!\omega_e &\propto e^{\frac{c_4}{2} BC} e^{\frac{1 + BC}{2} [c_1 (A+B) + c_2 AB]} e^{-\frac{1 - BC}{2} c_3 AB} = \exp \left[ \Omega_0 (A + B + C + ABC) + \Omega_{AB} AB + \Omega_{BC} BC + \Omega_{CA} CA \right],
        \end{align}
        where $\Omega_0 = c_1 / 2$, $\Omega_{AB} = (c_2 - c_3) / 2$, $\Omega_{BC} = c_4 / 2$, and $\Omega_{CA} = (c_2 + c_3) / 2$. Notice that $\Omega_0$ is always finite, while $\Omega_{AB}, \Omega_{CA} \sim c_2 / 2$ and $\Omega_{BC} \sim -c_2 / 2$ as we take $c_2 \rightarrow \infty$. Therefore, for all $0 < \gamma < 1$, the phase of the model is determined by the competition between the diverging coupling constants $\Omega_{AB}$, $\Omega_{BC}$, and $\Omega_{CA}$ and the effect of $\Omega_0$ is negligible. (The diverging coupling constants cannot cancel off each other.) Thus, neglecting $\Omega_0$ in Eq.~\eqref{weAmpDamp3} should not affect the phase of the model. Since the last term in Eq.~\eqref{weAmpDamp2} cannot appear when there is no $\Omega_0(A + B + C + ABC)$ term in Eq.~\eqref{weAmpDamp3}, we can conclude that we can safely drop the last term in Eq.~\eqref{weAmpDamp2} without altering the phase of the model. It then follows that
        \begin{align}
            \omega_e \propto 1 + \frac{\gamma (1 + \gamma)}{\gamma^2 - \gamma + 2} CA - \frac{\gamma (1 - \gamma)}{\gamma^2 - \gamma + 2} BC + \frac{\gamma^2 - 3\gamma + 2}{\gamma^2 - \gamma + 2} AB,
        \end{align}
        which is the weight $\omega_e \propto 1 + J_+ s_v s_{v'} - J_- \tau_v \tau_{v'} + K s_v s_{v'} \tau_v \tau_{v'}$ with $J_\pm$ and $K$ given by Eq.~\eqref{JK_AmpDamp} upon defining new Ising variables $s_v = x_v \bar{x}_v$ and $\tau_v = \bar{x}_v t_v$. Expressing this in the form of $\omega_e \propto e^{\JJ_1  s_v s_{v'} + \JJ_2 \tau_v \tau_{v'} + \KK s_v s_{v'} \tau_v \tau_{v'}}$, it must be
        \begin{equation} \label{TanhEqs}
            \begin{aligned}
                \frac{\tanh(\JJ_1) + \tanh(\JJ_2) \tanh(\KK)}{1 + \tanh(\JJ_1) \tanh(\JJ_2) \tanh(\KK)} &= J_+, \\
                \frac{\tanh(\JJ_2) + \tanh(\JJ_1) \tanh(\KK)}{1 + \tanh(\JJ_1) \tanh(\JJ_2) \tanh(\KK)} &= J_-, \\
                \frac{\tanh(K) + \tanh(\JJ_1) \tanh(\JJ_2)}{1 + \tanh(\JJ_1) \tanh(\JJ_2) \tanh(\KK)} &= K.
            \end{aligned}
        \end{equation}
        We have numerically checked that for $\gamma \in (0,1)$, the solutions of Eq.~\eqref{TanhEqs} are all finite reals. For $\gamma = 0$ ($1$), one gets $\JJ_1 = \JJ_2 = 0$ and $\KK = \infty$ ($\JJ_1 = \infty$ and $\JJ_2 = \KK = 0$). Therefore, the mixed-state phase of the TC under amplitude damping noise is associated with the anisotropic AT model (with $\JJ_2 \leq 0$).
    \end{widetext}

    \section{Monotonicity of R\'enyi Phase Boundaries} \label{App:Monotonicity}

    Strictly speaking, no theorem with full generality relates the R\'enyi entropic quantities to the replica-limit counterpart ($n = 1$). Nevertheless, we explain here why the R\'enyi-2 phase boundary $p_c^{(2)}$ should upper-bound the intrinsic threshold $p_c$ (i.e., $p_c \leq p_c^{(2)}$) under reasonable physical assumptions on the spectrum of the decohered density matrix.
    
\begin{figure}[t]
        \centering
        \includegraphics[width=0.83\columnwidth]{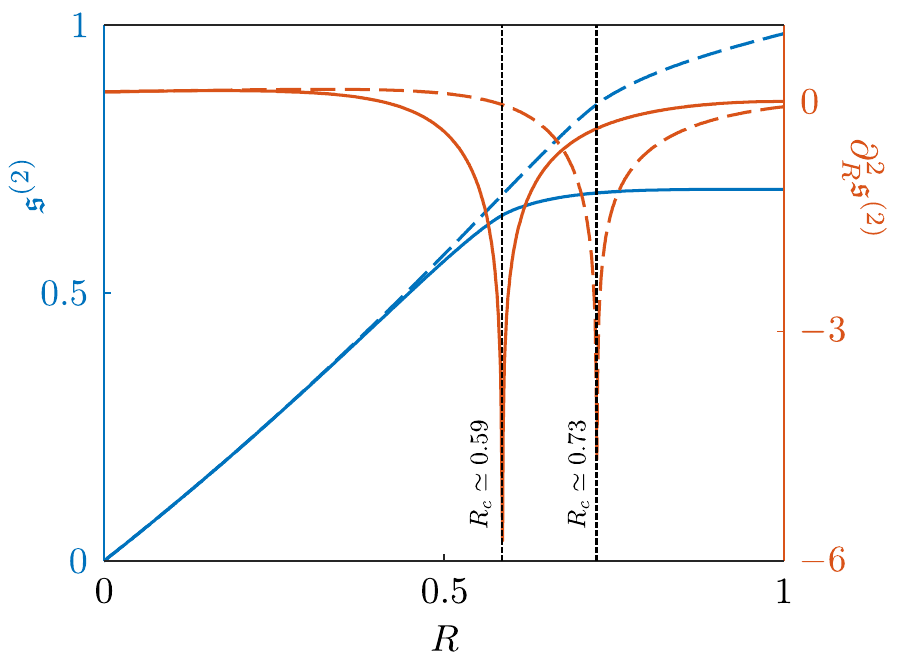}
        \caption{R\'enyi-2 entropy density $\mathfrak{s}^{(2)} \equiv \lim_{N\rightarrow \infty} S^{(2)} / N$ and its second derivative $\partial_R^2 \mathfrak{s}^{(2)}$ for the toric code under random rotation noise with $\phi = \theta = \pi/2$ (blue) and $\phi = \theta = 3\pi/8$ (orange). Divergences in $\partial_R^2 \mathfrak{s}^{(2)}$ at $R_c \simeq 0.59$ and $0.73$ mark the transition points. Data were obtained via the CTMRG algorithm with bond dimension $D = 50$.}
        \label{fig:2ndorder}
    \end{figure}

    Let $\rho_D(p)$ be a decohered density matrix at noise strength $p$, and $\lambda_j(p)$ its $j$th largest eigenvalue. Following a large-deviation principle, a spectrum of $\rho_D$ in the thermodynamic limit ($N\rightarrow \infty$) can be described by a variable $\varepsilon \geq 0$ and a rate function $s(\varepsilon, p)$, such that the number of eigenvalues $\lambda_j(p)$ in the window $[e^{-N\varepsilon}, e^{-N(\varepsilon + d\varepsilon)}]$ is $e^{N s(\varepsilon, p)} d\varepsilon$. In other words, $s(\varepsilon, p)$ plays the role of a ``microcanonical entropy density'' for eigenvalues at ``energy'' $\varepsilon$. In the large-$N$ limit, the R\'enyi-$n$ entropy density defined as $\mathfrak{s}^{(n)} (p) \equiv S^{(n)} (p) / N$ can be written as
    \begin{align}
        \mathfrak{s}^{(n)} (p) \simeq \frac{1}{1-n} \frac 1N \log \left[ \int_0^\infty d\varepsilon\, e^{N \Phi_n (\varepsilon, p)} \right],
    \end{align}
    where 
    \begin{align}
        \Phi_n (\varepsilon, p) \equiv s(\varepsilon, p) - n \varepsilon.
    \end{align}
    Applying the saddle-point approximation, we obtain
    \begin{align} \label{Renyidensity}
        \mathfrak{s}^{(n)} (p) \,\underset{N\rightarrow \infty}{\simeq}\, \frac{1}{1-n} \max_{\varepsilon \geq 0} \Phi_n (\varepsilon, p)
    \end{align}
    for all $n \in (0, 1) \cup (1, \infty)$. In the limit $n \rightarrow 1$, the R\'enyi entropy reduces to the von Neumann entropy, and one finds $\mathfrak{s}^{(1)}(p) = -\partial_n \Phi_n (\varepsilon, p) |_{n = 1}$. Therefore, $\mathfrak{s}^{(n)} (p)$ develops a nonanalyticity at a critical point $p = p_c^{(n)}$ where $\max_{\varepsilon \geq 0} \Phi_n (\varepsilon, p)$ becomes nonanalytic. Below, we focus on the second-order transition in $\mathfrak{s}^{(n)} (p)$, which is typical in decohered quantum memories (e.g., see Fig.~\ref{fig:2ndorder} for diverging $\partial_R^2 \mathfrak{s}^{(n)}$ under our random rotation noise). 
    
    \begin{figure}[h!]
        \centering
        \includegraphics[width=\columnwidth]{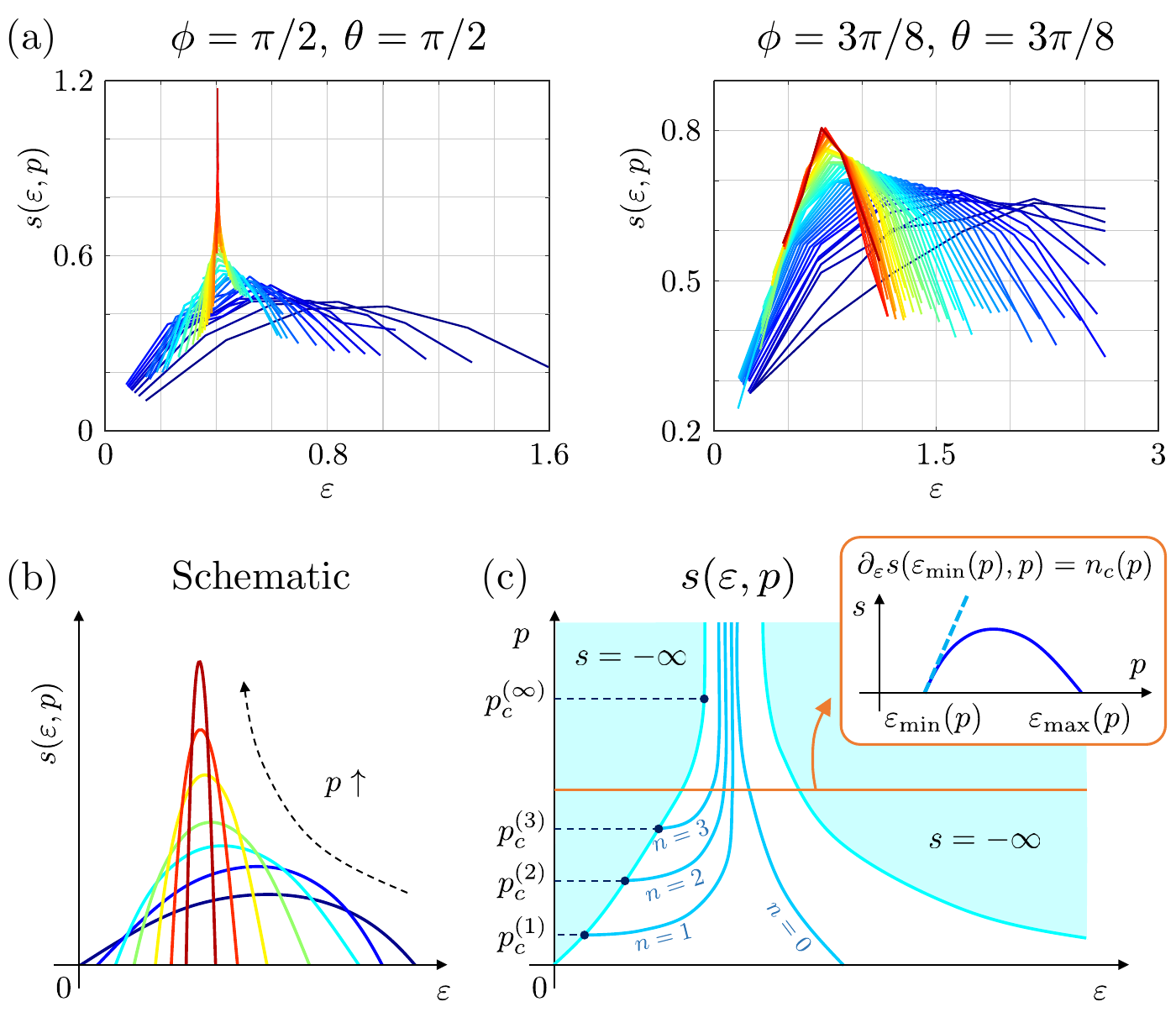}
        \caption{Estimated rate functions $s(\varepsilon, p)$ for the eigenvalues of decohered toric codes on a $3\times 2$ square lattice ($N = 12$ qubits) under random rotation noise with $\phi = \theta = \pi/2$ and $\phi = \theta = 3\pi/8$. Colors indicate noise strengths $p$ (blue for small $p$, red for large $p$; six bins were used per estimation. (b) Schematic of $s(\varepsilon, p)$ for the decohered toric codes: $s(\varepsilon, p)$ becomes sharper and narrower as $p$ increases. (c) General structure of $s(\varepsilon, p)$ in the ($\varepsilon$, $p$)-plane. Blue-shaded region corresponds to $s(\varepsilon, p) = -\infty$, and blue curves indicate interior saddle points for various $n$. The orange box depicts $s(\varepsilon, p)$ along the orange horizontal slice at fixed $p$.}
        \label{fig:rateftn}
    \end{figure}
    
    Based on physical intuition, we assume that the rate function $s(\varepsilon, p)$ for the eigenvalues of $\rho_D$ is of class $C^1$ and has the schematic shape shown in Fig.~\ref{fig:rateftn}(b). For any fixed $p$, $s(\varepsilon, p)$ is concave in $\varepsilon$ because it is the Legendre-Fenchel transform of $\log \mathrm{Tr} [\rho_D^n (p)]$, which is convex in $\varepsilon$. Moreover, since decoherence tends to flatten the spectrum of the density matrix, the interval $\varepsilon \in [\varepsilon_{\mathrm{min}}(p), \varepsilon_{\mathrm{max}}(p)]$ on which $s(\varepsilon, p) \geq 0$ shrinks as $p$ increases; along these boundaries, we have $s(\varepsilon_{\mathrm{min/max}}(p), p) = 0$, while $s(\varepsilon, p) = -\infty$ outside the interval, indicating that no eigenvalues occur there. Note also that the $\varepsilon$-slope of $s(\varepsilon, p)$ at $\varepsilon_{\mathrm{min}}(p)$ becomes steeper as $p$ increases, i.e., $\partial_p \partial_\varepsilon s(\varepsilon_{\mathrm{min}}(p), p) > 0$.  To readily understand the assumption, imagine a density matrix subject to depolarizing noise: the noise depletes eigenvalues that are too large or too small, and ultimately the state becomes the completely mixed state $2^{-N} I$, whose spectrum is perfectly flat. We have numerically estimated $s(\varepsilon, p)$ using noisy finite-size toric codes and confirmed that these assumptions hold well [see Fig.~\ref{fig:rateftn}(a)]. [Due to the limited system size, we reduced the number of bins to more accurately capture the overall shape of $s(\varepsilon, p)$.]
    
    In Eq.~\eqref{Renyidensity}, $\Phi_n (\varepsilon, p)$ can be maximized by two types of $\varepsilon$:
    \begin{enumerate}
        \item \emph{Interior saddle point} $\varepsilon_n^* (p) \in (\varepsilon_{\mathrm{min}}(p), \varepsilon_{\mathrm{max}}(p))$ satisfying
        \begin{align}
            \partial_\varepsilon s(\varepsilon_n^*(p), p) = n.
        \end{align}
        Since $s(\varepsilon, p)$ is concave in $\varepsilon$, such an interior solution can appear when the slope at the left boundary $\varepsilon_{\mathrm{min}}(p)$ exceeds $n$, i.e., $\partial_\varepsilon s(\varepsilon_{\mathrm{min}}(p), p) > n$. Defining $n_c(p) \equiv \partial_\varepsilon s(\varepsilon_{\mathrm{min}}(p), p)$ (which is positive for $p>0$) gives the critical $n$-value above which an interior saddle point can appear. 
    
        \item Otherwise (i.e., $n > n_c(p)$), the \emph{boundary point} $\varepsilon_{\mathrm{min}}(p)$ achieves the maximum of $\Phi_n (\varepsilon, p)$.
    \end{enumerate}
    Equivalently, for each fixed $n$ we define the R\'enyi-$n$ critical error rate $p_c^{(n)}$ by the relation 
    \begin{align} \label{pc_def}
        n = n_c(p_c^{(n)}) = \partial_\varepsilon s(\varepsilon_{\mathrm{min}}(p_c^{(n)}), p_c^{(n)}).
    \end{align}
    See Fig.~\ref{fig:rateftn}(c): the interior saddle curves $\varepsilon_n^* (p)$ (blue curves) meet the left boundary $\varepsilon_{\mathrm{min}}(p)$ at $p = p_c^{(n)}$. Since $s(\varepsilon_{\mathrm{min}}(p), p)$ is $C^1$, its first derivatives are continuous. However, since two different branches, $\varepsilon_{\mathrm{min}}(p)$ and $\varepsilon_n^* (p)$, meet at $p = p_c^{(n)}$, the second derivative $\partial_p^2 \mathfrak{s}^{(n)} (p)$ can be discontinuous, producing a second-order transition in $\mathfrak{s}^{(n)} (p)$. If the left boundary becomes vertical for high $p$, then no finite $n$ admits an interior saddle point. In this case, a possible second-order transition may arise from discontinuous second derivatives at the point $p = p_c^{(\infty)}$ where the left boundary becomes vertical.
    
    Now, we discuss the monotonicity of the transition points $p_c^{(n)}$ with respect to $n$. Figure~\ref{fig:rateftn}(c) already clearly shows that $p_c^{(n)}$ increases with $n$, but here we provide a formal proof. Differentiating Eq.~\eqref{pc_def} with respect to $n$ gives
    \begin{equation}
        \begin{aligned}
            1 &= \partial_p \partial_\varepsilon s(\varepsilon_{\mathrm{min}}(p), p) \frac{d p^{(n)}}{dn} + \partial_\varepsilon^2 s(\varepsilon_{\mathrm{min}}(p), p) \frac{d \varepsilon_{\mathrm{min}}(p)}{dn} \\
            &= \left[ \partial_p \partial_\varepsilon s(\varepsilon_{\mathrm{min}}(p), p) + \partial_\varepsilon^2 s(\varepsilon_{\mathrm{min}}(p), p) \frac{d\varepsilon_{\mathrm{min}} (p)}{d p} \right] \frac{d p^{(n)}}{dn}.
        \end{aligned}
    \end{equation}
    where all instances of $p$ evaluated at $p = p_c^{(n)}$. By concavity in $\varepsilon$, we have $\partial_\varepsilon^2 s(\varepsilon_{\mathrm{min}}(p_c^{(n)}), p_c^{(n)}) > 0$. Also, differentiating $s(\varepsilon_{\mathrm{min}}(p), p) = 0$ with respect to $p$ yields
    \begin{align}
        \partial_\varepsilon s(\varepsilon_{\mathrm{min}}(p), p) \cdot \frac{d \varepsilon_{\mathrm{min}}(p)}{dp} + \partial_p s(\varepsilon_{\mathrm{min}}, p) = 0, 
    \end{align}
    so at $p = p_c^{(n)}$, we obtain
    \begin{align}
        \frac{d\varepsilon_{\mathrm{min}} (p_c^{(n)})}{d p} = -\frac{\partial_p s(\varepsilon_{\mathrm{min}} (p_c^{(n)}), p_c^{(n)})}{n_c(p)},
    \end{align}
    which is positive since $\partial_p s(\varepsilon_{\mathrm{min}} (p_c^{(n)}), p_c^{(n)}) < 0$ [which is clear from the assumed form of $s(\varepsilon, p)$ in Fig.~\ref{fig:rateftn}(b)] and $n_c(p) > 0$ . Combining with $\partial_p \partial_\varepsilon s(\varepsilon_{\mathrm{min}}(p_c^{(n)}), p_c^{(n)}) > 0$, we conclude that $d p_c^{(n)} / dn > 0$, i.e., $p_c^{(n)}$ is an increasing function of $n$. Therefore,
    \begin{align} \label{monotone_pc}
        p_c^{(n_1)} \leq p_c^{(n_2)} \quad\text{for}\quad n_1 < n_2,
    \end{align}
    given our physical spectral assumptions on $\rho_D$. This argument holds for all $n \in (0, 1) \cup (1, \infty)$, so we expect Eq.~\eqref{monotone_pc} to extend to the $n = 1$ case, in particular, $p_c^{(1)} \leq p_c^{(2)}$. To our knowledge, all previously studied mixed-state quantum many-body systems obey this monotonicity in the R\'enyi index $n$. We expect other R\'enyi-$n$ information-theoretic quantities of $\rho_D$ to undergo a transition at the same error rate $p = p_c^{(n)}$, since they too can be expressed in terms of $s(\varepsilon, p)$ and $n$, or viewed as observables in the same statistical-mechanics model that governs the $n$th moment $\mathrm{Tr}[\rho_D^n]$.
    
    For cases with special spectral symmetries in noise parameters, the above argument requires only a minor adjustment. For instance, the toric code ground state $|\Psi_0\rangle$ (with $\mathbf{Z}_{1,2} |\psi_0\rangle = |\Psi_0\rangle$) under amplitude damping noise with strength $\gamma$ has the same spectrum as for damping strength $1 - \gamma$. This spectral symmetry implies that the rate function $s(\varepsilon, \gamma)$ is symmetric about $\gamma = 1/2$. Consequently, the first mixed-state transition point $\gamma_{c,1}^{(n)}$ (from the quantum-memory phase to the classical-memory phase) increases with the R\'enyi index $n$, while the second transition point $\gamma_{c,2}^{(n)} = 1 - \gamma_{c,1}^{(n)}$ decreases with $n$, preserving reflection symmetry about $\gamma = 1/2$. (Imagine reflecting Fig.~\ref{fig:rateftn} vertically about $\gamma = 1/2$.) 
\end{appendix}

\bibliography{ref}

\end{document}